\documentclass[manuscript,screen,final]{acmart}
\AtBeginDocument{%
  }

\usepackage{amssymb}
\usepackage{amsmath}
\usepackage{amsmath,amsfonts}
\usepackage{multirow} 
\usepackage{algorithmic}
\usepackage{algorithm}
\usepackage{array}
\usepackage[caption=false,font=normalsize,labelfont=sf,textfont=sf]{subfig}
\usepackage{textcomp}
\usepackage{stfloats}
\usepackage{url}
\usepackage{verbatim}
\usepackage{graphicx}
\usepackage{orcidlink}
\usepackage{colortbl}
\usepackage{breqn}
\usepackage{textcomp}
\usepackage{amsmath,amsfonts}
\usepackage{algorithm}
\usepackage{algorithmic}
\usepackage{graphicx}
\usepackage{textcomp}
\usepackage{xcolor}
\usepackage{longtable}
\usepackage{tcolorbox}
\usepackage{multirow}
\usepackage[normalem]{ulem}
\usepackage{url}
\usepackage{xcolor}
\usepackage{tikz}
\usepackage{mdframed}
\newcommand{\blackcircled}[1]{%
  \tikz[baseline=(char.base)]{
    \node[shape=circle,fill=black,inner sep=0.3pt] (char)
    {\textcolor{white}{\sffamily\bfseries #1}};}}
\newcommand{\cmtwidth}{.40\linewidth}

\newcommand{\rcmt}[1]{%
  \hfill
  \parbox[t]{\cmtwidth}{\raggedleft$\triangleright$~#1}%
}

\begin{document}

\title{BASFuzz: Towards Robustness Evaluation of LLM-based NLP Software via Automated Fuzz Testing}

\author{Mingxuan Xiao}
\email{xiaomx@hhu.edu.cn}
\orcid{0009-0008-2800-3306}
\affiliation{%
  \institution{Hohai University}
  \city{Nanjing}
  \country{China}
}

\author{Yan Xiao}
\authornote{Corresponding author}
\affiliation{%
  \institution{Sun Yat-sen University}
  \city{Shenzhen}
  \country{China}}
\email{xiaoy367@mail.sysu.edu.cn}
\orcid{0000-0002-2563-083X}

\author{Shunhui Ji}
\affiliation{%
  \institution{Hohai University}
  \city{Nanjing}
  \country{China}
}
\orcid{0000-0002-8584-5795}
\email{shunhuiji@hhu.edu.cn}

\author{Jiahe Tu}
\affiliation{%
  \institution{Hohai University}
  \city{Nanjing}
  \country{China}
}
\orcid{0009-0001-3604-8378}
\email{tujh@hhu.edu.cn}

\author{Pengcheng Zhang}
\affiliation{%
  \institution{Hohai University}
  \city{Nanjing}
  \country{China}
}
\authornotemark[1]
\orcid{0000-0003-3594-408X}
\email{pchzhang@hhu.edu.cn}



\renewcommand{\shortauthors}{M. Xiao et al.}

\begin{abstract}
Fuzzing has shown great success in evaluating the robustness of intelligent natural language processing (NLP) software. As large language model (LLM)-based NLP software is widely deployed in critical industries, existing methods still face two main challenges: \blackcircled{1} testing methods are insufficiently coupled with the behavioral patterns of LLM-based NLP software; \blackcircled{2} fuzzing capability for the testing scenario of natural language generation (NLG) generally degrades.
To address these issues, we propose BASFuzz, an efficient \underline{Fuzz} testing method tailored for LLM-based NLP software. BASFuzz targets complete test inputs composed of prompts and examples, and uses a text consistency metric to guide mutations of the fuzzing loop, aligning with the behavioral patterns of LLM-based NLP software. A \underline{B}eam-\underline{A}nnealing \underline{S}earch algorithm, which integrates beam search and simulated annealing, is employed to design an efficient fuzzing loop. In addition, information entropy-based adaptive adjustment and an elitism strategy further enhance fuzzing capability. We evaluate BASFuzz on six datasets in representative scenarios of NLG and natural language understanding (NLU). Experimental results demonstrate that BASFuzz achieves a testing effectiveness of 90.335\% while reducing the average time overhead by 2,163.852 seconds compared to the current best baseline, enabling more effective robustness evaluation prior to software deployment.
\end{abstract}




\maketitle

\section{Introduction}
\begin{figure}[t]
    \centering
    \includegraphics[width=0.9\textwidth]{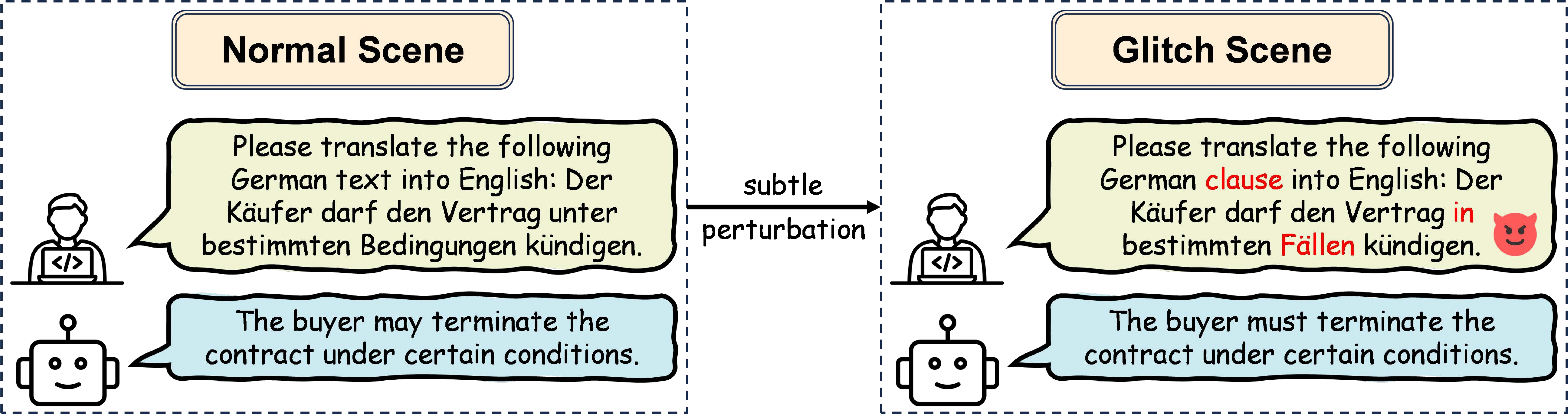}\\
    \caption{An example of robustness flaws in LLM-based NLP software. In the contract translation scenario, slight perturbations (red) in the prompt and example section cause the software to mistranslate a clause expressing permission to terminate the contract into an obligation to terminate.} 
    \label{Fig1}
\end{figure}

\textbf{Large language model (LLM)}-based software, as a crucial component of intelligent software, refers to software systems that utilize LLMs as core components to perform language understanding, generation, or reasoning tasks~\cite{10.1145/3719006,11029790}. With the remarkable progress of LLMs in numerous \textbf{natural language processing (NLP)} tasks, LLM-based NLP software has gradually evolved from research prototypes into deployable product systems. It has been widely applied in practical scenarios such as machine translation software~\cite{10.1145/3664608}, intelligent question-answering assistants~\cite{10.1145/3730578}, and financial analysis systems~\cite{10.1145/3688399}. However, similar to traditional intelligent software, concerns have been raised regarding the robustness of LLM-based NLP software~\cite{10.1145/3745764}. Specifically, when confronted with carefully crafted inputs containing subtle perturbations, the software's output often exhibits highly inconsistent behaviour, resulting in semantic misjudgments or logical errors, thereby jeopardizing software reliability and user trust. As shown in Figure~\ref{Fig1}, in a contract translation scenario, the translation software, due solely to subtle wording mutations in the input, mistranslated a clause expressing permission to terminate a contract into an obligation to terminate, causing a significant semantic error. Such behavioural instability not only affects user experience but may also trigger catastrophic consequences in high-stakes domains such as law~\cite{harasta2024cannot} and healthcare~\cite{qiu2024llm}, including invalid contracts and ambiguous attribution of breach liabilities. Therefore, testing LLM-based NLP software to evaluate its robustness is of critical importance~\cite{10.1145/3631977,10.1145/3716167}.

In the software development process, testing has always been regarded as a critical means to ensure software quality and uncover potential flaws. Current work primarily focuses on jailbreak robustness~\cite{10.1145/3724393,10.1145/3691620.3695001}, fairness robustness~\cite{10.1145/3737697}, and factual robustness~\cite{10.1145/3744340}. These methods are respectively used to evaluate whether LLM-based software can resist toxic prompting, maintain neutrality towards sensitive attributes (e.g., gender, race), and preserve factual consistency and accuracy. Although these directions are of significant value in ensuring software reliability and security, \textbf{adversarial robustness}, as a form of robustness more closely aligned with real-world applications, remains insufficiently explored in current research~\cite{10.1145/3719006,10.1145/3731559}, with this gap being particularly evident in \textbf{natural language generation (NLG)} tasks. Unlike \textbf{natural language understanding (NLU)} tasks such as text classification or semantic entailment, user inputs in NLG tasks are open-ended and compositional natural language texts, making them highly susceptible to adversarial perturbations~\cite{10.1145/3660808}. For example, input perturbations similar to those shown in Figure~\ref{Fig1} may not necessarily originate from malicious attackers. However, they could also arise from challenging or borderline content unintentionally created by users during actual usage, leading the software to exhibit severely inconsistent behaviour. More critically, such flaws are often only exposed through user feedback after software deployment, resulting in significantly increased costs for fault remediation and rollback~\cite{10.1145/3714463}. Therefore, there is an urgent need for a pre-deployment adversarial robustness testing method to evaluate the robustness of LLM-based NLP software in NLG tasks.

Fuzz testing is a widely adopted automated testing technique. Its core principle is to automatically generate a large number of mutated input examples to explore the boundaries of software behaviour and identify hidden vulnerabilities~\cite{8863940}. This approach has proven effective in security testing~\cite{10.1145/3744242} and vulnerability discovery~\cite{10.1145/3580596}. In recent years, with the rapid adoption of intelligent software, researchers have extended fuzz testing to deep learning models~\cite{10172506}. Target systems include image classifiers~\cite{10.1145/3688835}, speech recognition systems~\cite{9540914}, and text classification models~\cite{10298415}. These studies typically introduce small input perturbations or guide model decision paths to expose the inconsistent behaviours of neural networks (named threat models) under specific conditions, thereby enabling robustness evaluation. Existing fuzzers can be classified into generation-based, learning-based, and mutation-based approaches~\cite{10.1145/3597503.3639121}. Generation-based fuzzers create input data offline using predefined mutation rules. Learning-based fuzzers employ neural networks to generate test cases. Mutation-based fuzzers iteratively execute fuzzing loops following mutation rules, often combined with search-based software engineering techniques~\cite{10.1145/3715002}. However, all three fuzzing approaches face the following challenges when used in testing LLM-based NLP software:

\textbf{Challenge \#1: Insufficient coupling between testing methods and the behavioural patterns of LLM-based NLP software.}
Most existing robustness testing methods for LLM-based NLP software focus on NLU tasks such as text classification~\cite{10903339} and sentiment analysis~\cite{xiao2024assessing}. These tasks have relatively fixed output structures and limited label sets. Their primary goal is to evaluate the software's sensitivity under structured output settings, making it easier to identify the robustness flaws caused by perturbed inputs. In practical applications, however, users more frequently interact with LLM-based software through NLG tasks, including machine translation~\cite{10.1145/3664608} and dialogue generation~\cite{10.1145/3691620.3695018}. These tasks exhibit highly open input contexts, diverse output spaces, and more complex control over linguistic styles, which makes conventional testing approaches inadequate for capturing robustness defects~\cite{10.1145/3719006,10.1145/3731559}. As illustrated in Figure~\ref{Fig1}, the software produces semantically directional mistranslations when faced with subtle linguistic perturbations (e.g., ``unter Bedingungen’’ vs. ``in Fällen’’), highlighting the unique challenges of adversarial robustness in NLG tasks. Furthermore, the multilingual nature of tasks such as translation amplifies this problem, requiring testing methods with stronger semantic modelling and cross-linguistic adaptation capabilities.

\textbf{Challenge \#2: Degraded fuzzing capability in the testing scenario of NLG.}
We evaluate the performance of existing fuzzing techniques~\cite{BASFuzz} when applied to the NLG scenario. The evaluation includes fuzzers designed for LLM-based NLU tasks~\cite{10.24963/ijcai.2024/730,xiao2025abfs,xiao2024assessing} and those developed for traditional \textbf{deep neural network (DNN)}-based NLG tasks~\cite{Cheng_Yi_Chen_Zhang_Hsieh_2020,tan-etal-2020-morphin,Jin_Jin_Zhou_Szolovits_2020,ren-etal-2019-generating}.
The former shows a reduction of 14.198\%–85.191\% in testing effectiveness on generative tasks,
primarily due to their inability to adapt to open-ended output spaces. The latter have shown success in generative settings; however, when transferred to LLM-based software, their testing efficiency is severely constrained and fails to capture the dynamic response behaviours of threat models. These findings suggest that when encountering LLM-based NLP software with strong generalisation capabilities and complex language generation patterns~\cite{10.1145/3715007}, even after unifying and refining the objective function, the original mutation strategies and sampling logic are insufficient to expose robustness flaws. Therefore, it is essential to design an efficient fuzzing approach with stronger task adaptability for NLG tasks, enabling effective robustness evaluation in open-ended generative scenarios.

\textbf{Proposed Technique.} To address the above challenges, this paper proposes BASFuzz, a fuzzing approach for evaluating the adversarial robustness of LLM-based NLP software. To align with real-world software usage scenarios, BASFuzz adopts a holistic input perturbation strategy, treating user prompts and task examples as a unified testing unit. Considering the open-ended nature of output spaces in generative tasks, BASFuzz employs the \textbf{bilingual evaluation understudy (BLEU)} score~\cite{10.3115/1073083.1073135} as a text consistency metric to guide fuzzing towards perturbations that are minimal in input mutation but impactful on output, thereby mitigating the coupling gap identified in Challenge \#1. To ensure that perturbed test cases remain semantically aligned with the original input, BASFuzz introduces a mutation strategy based on multilingual word network~\cite{81a98633248a44d6864e3906cba8c3b5} and LLM-based word embeddings~\cite{DBLP:journals/corr/abs-2409-15700}. By combining perturbations with semantic constraints, it constructs a diverse set of variants, making it particularly effective for robustness evaluation in multilingual NLG tasks. During iterative exploration of the perturbation space, conventional search methods often fall into local optima due to the high-dimensional input space of LLM-based software. To mitigate this, BASFuzz integrates beam search~\cite{lowerre1976harpy} with a simulated annealing~\cite{doi:10.1126/science.220.4598.671} mechanism to form a dynamic search process with global exploration capability. Beam search maintains multiple perturbation paths in parallel to improve coverage of the input space. At the same time, simulated annealing allows the acceptance of perturbation paths with slight quality degradation to escape early search traps, enabling the discovery of deeper robustness. Considering the constraint of computational resources in practical testing, BASFuzz incorporates the information entropy~\cite{6773024} of candidate perturbation distributions as a dynamic signal to adjust the beam width in real-time. In high-entropy phases, the search expands to enhance exploration; in low-entropy phases, it converges to improve efficiency. This adaptive mechanism achieves efficient testing while enhancing the effectiveness of robustness evaluation.

We select five threat models with varying parameter scales and three real-world machine translation datasets to evaluate BASFuzz's effectiveness in multilingual settings. The baselines include six of the most recent and representative robustness testing approaches to ensure broad comparability of results. Experimental results show that existing methods are insufficient for effectively testing LLM-based NLP software. For example, VFA achieves a testing success rate of 57.958\% on Llama3\_70B, whereas BASFuzz reaches 89.138\%, indicating its ability to uncover significantly more robustness flaws during the software testing phase. We further evaluate the quality of BASFuzz-generated test cases using change rate, perplexity, and grammatical error count. Across 15 experimental settings, BASFuzz achieves an average change rate of only 3.748\%, with both perplexity and grammatical errors lower than those of existing baselines. This demonstrates that BASFuzz produces more covert input perturbations and generates more fluent test cases. BASFuzz not only efficiently exposes robustness flaws in LLM-based NLP software across different languages and threat model scales, but also demonstrates strong test transferability and task scalability. The main \textbf{contributions} of this work are as follows:
\begin{itemize}
\item \emph{Method.} We propose BASFuzz, a mutation-based fuzzing method designed for LLM-based NLP software to evaluate adversarial robustness in NLG tasks. BASFuzz supports black-box testing and is built to address the weak mutation capability and low testing effectiveness of existing approaches in generative tasks. It combines a word network and LLM-based word embeddings to automatically generate multidimensional semantic perturbations. The input perturbation space is explored using a beam-annealing search algorithm with adaptive path maintenance and diversity guidance, enhancing the evaluation of software robustness.

\item \emph{Tool.} We implement and release an open-source mutation-based fuzzing tool~\cite{BASFuzz} that provides full automation, covering objective function selection, input perturbation generation, and search control. The tool provides a unified interface and configuration mechanism, eliminating task-type restrictions inherent in existing testing frameworks. It supports multi-task extension: testing strategies originally designed for NLG tasks can be directly applied to NLU tasks and vice versa, significantly improving the scalability of current robustness testing methods.

\item \emph{Study.} We conduct a series of comparative and ablation studies on three multilingual translation datasets using five mainstream LLMs as threat models and six baseline methods. Experimental results demonstrate that BASFuzz outperforms existing approaches in terms of testing effectiveness, efficiency, and scalability. It demonstrates stronger robustness evaluation capability and better adaptability to practical applications, providing empirical evidence for advancing robustness testing in LLM-based NLP software engineering.
\end{itemize}
The remainder of this paper is organised as follows: Section~\ref{sec2} introduces adversarial robustness and conventional search strategies. Section~\ref{sec3} describes the research motivation for designing BASFuzz. Section~\ref{sec4} presents the BASFuzz testing method. Section~\ref{sec5} details the datasets, threat models, and six baselines used in the experiments. Section~\ref{sec6} evaluates the fuzzing capability of BASFuzz across six metrics. Section~\ref{sec7} discusses threats to validity. Section~\ref{sec8} reviews existing work on the robustness testing of NLP software. Finally, Section~\ref{sec9} concludes the paper.

\section{Background}\label{sec2}
\subsection{Robustness of LLM-based NLP software}
LLMs are emerging as general-purpose natural language processing engines and are increasingly becoming the core computational components of intelligent software. In recent years, LLM-based NLP software has been widely deployed in various safety-critical scenarios, such as traffic rule parsing in autonomous driving~\cite{10998949}, text classification in financial trading platforms~\cite{10.1145/3677052.3698696}, and question answering generation in medical decision-support systems~\cite{TAN2024108290}. Robustness issues are particularly prominent in LLM-based software. LLMs operate in high-dimensional embedding spaces and rely on prompt–example combinations to model semantics, making them extremely sensitive to the overall input~\cite{xiao2024assessing,yu-etal-2024-freeeval}. Compared to traditional DNNs, LLMs contain orders of magnitude more parameters. They are trained on far more complex data, which amplifies their sensitivity to input details and further increases behavioural uncertainty. Prior studies have shown that LLM-based software often produces entirely different outputs for inputs with slight mutations in expression. This phenomenon is especially pronounced in generative tasks, posing a direct threat to software reliability and user trust.

According to the IEEE definition~\cite{iso2017iso}, robustness in software engineering is ``degree to which a system, product or component performs specified functions under specified conditions for a specified period of time’’. In the machine translation software, let the threat model $\mathcal{F}$ be trained on $(T_x, T_y) \sim \mathcal{D}$, where $T_x$ is the source-language text and $T_y$ is the corresponding reference translation. A practical user input can be formed by concatenating a prompt template $P$ with $T_x$, denoted as $I = \left[P; T_x\right]$. We focus on the following question: for perturbed $P^{\prime}$ and $(T_x^{\prime}, T_y)\sim \mathcal{D}^{\prime}\neq \mathcal{D}$, yielding an adversarial test case $I_{adv}=[P^{\prime}; T_x^{\prime}]$, can the software generate a translation result that remains as consistent as possible with $T_y$? If the software frequently produces outputs significantly different from $T_y$ on $P^{\prime}$ and $\mathcal{D}^{\prime}$, it indicates a lack of necessary robustness when faced with natural language mutations. 
Therefore, the above robustness testing problem can be formalized as finding an adversarial input $I_{adv}$ within the perturbation space $C(I_{ori})$, such that subtle perturbations can lead to changes in the software output~\cite{szegedy2014intriguing}:
\begin{equation}
\begin{aligned}
& \underset{I_{a d v} \in C\left(I_{o r i}\right)}{\textit{arg\,min} } \Delta (I_{o r i}, I_{a d v}) \\
& \text { s.t. } \mathcal{F}\left(I_{o r i}\right) \neq \mathcal{F}\left(I_{a d v}\right)
\end{aligned}
\label{eq1}
\end{equation}
where $\Delta (I_{o r i}, I_{a d v})$ denotes the perturbation magnitude metric, such as word replacement rate or text perplexity, and C represents the perturbation constraints to ensure the stealthiness of test cases. With the widespread deployment of LLM-based NLP software in mission-critical domains, developing efficient and scalable testing methods has become a key step in evaluating software robustness.

\subsection{Beam Search}
Beam search is a heuristic search strategy widely used to approximate optimal solutions in high-dimensional discrete spaces, particularly in problems involving a combinatorial explosion with exponential-scale output spaces. The algorithm was first applied to speech recognition~\cite{lowerre1976harpy}, which has since been extended to natural language generation~\cite{10.1145/3539618.3591698} and code completion~\cite{zelikman2024selftaught}. Compared with single-path greedy search~\cite{dantzig1957discrete}, beam search retains multiple candidate paths at each step, striking a balance between accuracy and computational efficiency. This design provides greater search diversity and a stronger ability to avoid local optima.

The core idea of beam search is that, at each generation step $t$, the algorithm selects the top-$k$ candidate paths (where $k$ is the beam width) based on their cumulative scores and expands them before pruning. Taking machine translation as an example, let the current search state at step $t-1$ be the candidate set $\mathcal{B}_{t-1} = {y_1^{t-1}, \ldots, y_k^{t-1}}$, where the score of each path $y_i^{t-1}$ is denoted as $S(y_i^{t-1})$. The expansion and selection at step $t$ proceed as follows:

(1) Expansion: For each path $y_i^{t-1}$ in $\mathcal{B}_{t-1}$, generate extended paths $y_i^{t} = [y_i^{t-1}; w]$ according to the next-token probability distribution $p(\cdot|y_i^{t-1})$ predicted by the machine translation model.

(2) Scoring: Compute the cumulative score $S(y_i^{t})$ for each extended path, typically using a log-probability accumulation:
\begin{equation}
S(y_i^{t}) = \log p(w_1, \ldots, w_t) 
= \sum_{j=1}^{t} \log p(w_j \mid w_{<j})
\label{eq2}
\end{equation}

(3) Pruning: Select the top-$k$ paths with the highest scores from all extensions to form the new candidate set $\mathcal{B}_t$.

(4) Termination: When all candidate paths satisfy stopping conditions, such as containing an end-of-sequence token or reaching the maximum length, output the highest-scoring path as the final result.

Beam search imposes a width constraint at each step to control computational complexity and avoid the exponential cost of full-space search. Although beam search is effective in handling exponential search spaces, its use in robustness testing for LLM-based NLP software faces limitations. A fixed beam width often leads to premature convergence in the perturbation space. Search paths become dominated by the initial quality of mutations, increasing the risk of local optima. Moreover, fixed-width strategies fail to account for the dynamic complexity of the perturbation process, making it difficult to adapt resource allocation to the evolving search state. This limitation can result in computational redundancy or the omission of critical perturbation paths.

\subsection{Simulated Annealing}
Simulated annealing is a global optimization algorithm based on stochastic search, inspired by the thermodynamic behaviour of metal annealing in solid-state physics. First introduced by Kirkpatrick et al.~\cite{doi:10.1126/science.220.4598.671} in 1983, simulated annealing has been widely applied to combinatorial optimization problems, including circuit layout~\cite{9771178}, production scheduling~\cite{LIU2023257}, and feature selection~\cite{MOLLA2022149728}. In recent years, simulated annealing has also been adopted in software testing~\cite{ZHU2021111026} and deep learning~\cite{KASSAYMEH2022108511} to prevent search processes from being trapped in local optima and to enhance diversity in exploring the perturbation space.

Simulated annealing constructs a non-greedy search process with a temperature control mechanism. At high temperatures in the early search phase, it allows acceptance of inferior solutions to facilitate broad exploration of the solution space. As the temperature gradually decreases, the algorithm transitions to fine-grained local optimization, effectively mitigating the tendency of traditional greedy search to converge to local optima. Let the objective be to minimize a target function $\mathcal{L}(s)$, where $s$ denotes a state in the search space, such as a candidate test case. The standard simulated annealing procedure consists of:

(1) Initialization: Randomly select an initial state $s_0$ and set the initial temperature $tem_0$.

(2) Perturbation generation: Generate a candidate solution $s^{\prime}$ in the neighbourhood of the current state $s_t$.

(3) Energy evaluation: Compute the change in the objective function, $\Delta \mathcal{L} = \mathcal{L}(s^{\prime}) - \mathcal{L}(s_t)$.

(4) Probabilistic acceptance: Apply the Metropolis criterion~\cite{metropolis1953equation} to decide whether to accept $s^{\prime}$:
\begin{itemize}
\item If $\mathcal{L}(s^{\prime})<\mathcal{L}(s_t)$: always accept.
\item If $\mathcal{L}(s^{\prime}) \geq \mathcal{L}(s_t)
$: accept with probability $p = \exp(-\Delta \mathcal{L} / tem_t)
$, where $tem_t$ is the current temperature.
\end{itemize}

(5) Temperature update: Update the temperature using the annealing schedule $tem_t\rightarrow tem_{t+1}$.

(6) Termination: Stop when the maximum number of iterations is reached or the temperature drops below a predefined threshold.

A common temperature schedule is exponential decay, $tem_t = \gamma \cdot tem_{t-1}$, where $\gamma\in (0,1)$ is the cooling factor. As $tem_t$ decreases, the algorithm shifts from high-temperature global exploration with high freedom to low-temperature local convergence, achieving a controlled balance between exploration and exploitation. However, when applied to robustness testing, especially for NLG tasks in LLM-based software, simulated annealing exhibits two limitations. It performs perturbation jumps on a single path without a path management mechanism, making it prone to missing high-quality perturbation sequences. Additionally, fixed cooling schedules fail to adapt to the dynamic fluctuations of the perturbation process, thereby reducing testing efficiency. To overcome these limitations, this paper designs a beam-annealing search strategy (cf. Section~\ref{sec4_5}) that integrates multi-path management of beam search with logarithmic-decay simulated annealing, thus achieving a better balance between global exploration and testing efficiency.

\section{Motivation}\label{sec3}
The software testing phase inevitably incurs substantial labor and resource costs. In the United States alone, software testing labor expenses amount to approximately 48 billion USD annually~\cite{10.1145/3611643.3616327}. In domains such as legal contracts~\cite{harasta2024cannot} and financial analysis~\cite{10.1145/3688399}, constructing test cases and validating results for NLP software often require domain experts, further increasing testing investment. The high human and time costs of manual testing~\cite{10.1145/3699598} severely limit its feasibility for robustness evaluation of LLM-based software and make it prone to missing edge cases. Moreover, the input feature space of LLM-based software is not a finite discrete set in the traditional sense. Inputs are represented in a high-dimensional discrete word embedding space; even for short sentences in translation or classification tasks, the number of potential linguistic variants reaches billions. This immense combinatorial space makes it virtually impossible to manually craft test cases that exhaustively cover potential risk variants. Therefore, robustness testing for LLM-based NLP software must rely on automated testing to balance effectiveness and efficiency.

In terms of testing paradigms, white-box methods depend on access to internal gradients and network structures. However, in most real-world engineering scenarios, LLMs are provided as encapsulated APIs or commercial services, leaving developers without access to internal parameters or gradients. Even with open-source models, the sheer scale of LLM parameters makes efficient access during testing impractical~\cite{10.1145/3708528}. In contrast, black-box testing is not only a practical necessity but also inherently advantageous for supporting diverse testing scenarios and task generality. Black-box approaches evaluate robustness solely through input–output interactions, avoiding dependency on threat model internals. For these reasons, BASFuzz is designed to perform robustness testing under black-box settings.
It is worth noting that many current generation-based fuzzers~\cite{wang2024robustness, NEURIPS2023_b6b5f50a} adopt a static testing paradigm in which a large set of test cases is generated offline, then executed in batch against the threat model. This design introduces two critical issues. First, static test data quickly becomes obsolete as LLMs are updated, failing to capture the behavioral patterns of new model versions. Second, generating large-scale offline data carries a high risk of overlapping with training data, which may lead to data leakage and biased testing results. Learning-based fuzzers~\cite{10.1145/3688839, 10.1145/3597503.3639118, 10.1145/3713081.3731733} attempt to address this by training DNNs to learn input perturbation distributions. However, during testing, they must execute both the fuzzer and the target LLM simultaneously, which significantly increases runtime. Our empirical evaluation of popular learning-based fuzzers~\cite{li-etal-2020-bert-attack,garg-ramakrishnan-2020-bae} on LLM-based NLP software shows that generating a single valid test case requires more than 300 minutes. This result fundamentally contradicts the goal of automated testing, which is to reduce testing costs and accelerate software iteration.
BASFuzz belongs to the mutation-based fuzzing class~\cite{10.1145/3691620.3695001, 11029790, 10.1145/3688835, 10298415}. Such fuzzers perform an iterative search over candidate variants and dynamically adjust mutation strategies based on real-time feedback from the threat model. It does not rely on secondary training or offline data storage, thereby avoiding dataset obsolescence and enabling continuous exploration of the perturbation space. This capability significantly improves the detection of complex robustness flaws. For NLG tasks, where the output space is highly open-ended, mutation-based fuzzing establishes a closed-loop validation process between perturbations and software responses, making it more adaptive and practically valuable.

To address the high-dimensional and discrete nature of the perturbation space in LLM-based NLP software, we design a new hybrid heuristic search algorithm to perform iterative mutation fuzzing. In BASFuzz, beam search is introduced to address the combinatorial explosion of natural language inputs. Effective software testing requires broad coverage of such high-dimensional yet semantically sensitive spaces. Single-path greedy perturbation easily converges prematurely, missing effective test inputs. Beam search mitigates this by maintaining multiple perturbation paths in parallel at each iteration, delaying convergence decisions and improving both coverage and perturbation diversity, which satisfies the software engineering requirement for representative test cases. At the same time, considering the nonlinear and highly sensitive decision boundaries of LLM outputs, we integrate simulated annealing into the beam search process. Its core is a temperature-controlled probabilistic acceptance mechanism that allows degraded paths to be retained with a certain probability, thereby escaping local convergence traps. This design enables more global exploration of the perturbation space. It is particularly suited to robustness evaluation in black-box testing scenarios where gradient information and interpretable internal states are unavailable.

\section{Methodology}\label{sec4}
In this section, we elaborate on \textbf{BASFuzz}, our method for robustness testing of LLM-based NLP software. We first provide an overview of BASFuzz and then detail its key components.
\subsection{Overview of BASFuzz}
\begin{figure}[t]
    \centering
    \includegraphics[width=1\textwidth]{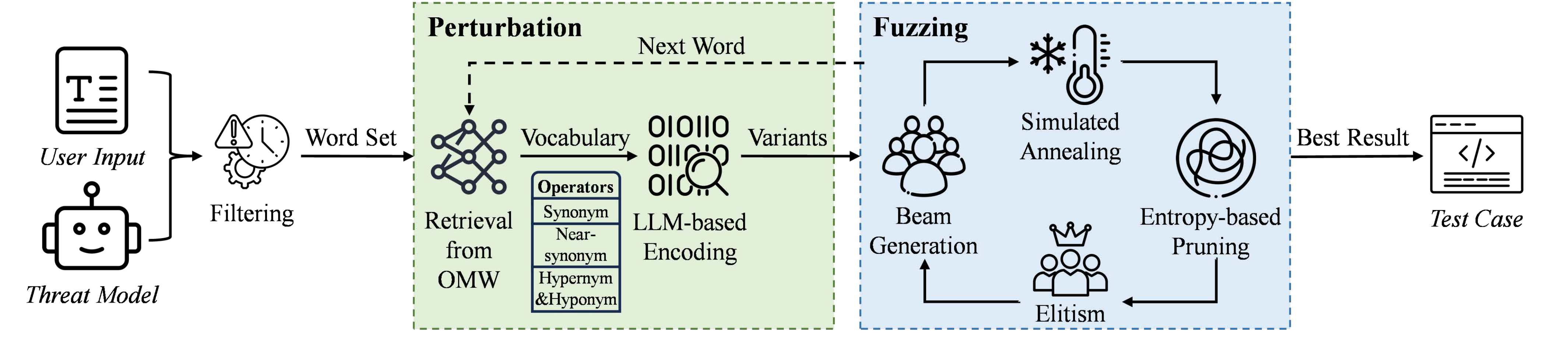}\\
    \caption{Overview of BASFuzz.} 
    \label{Fig2}
\end{figure}

To evaluate the robustness of LLM-based NLP software under black-box settings, we propose BASFuzz. This automated testing method combines LLM-driven perturbation space construction with a hybrid search-based fuzzing loop. The overall architecture of BASFuzz is shown in Figure~\ref{Fig2}. The entire testing process is guided by a robustness objective function (cf. Section~\ref{sec4_2}), and the fuzzing loop terminates when the objective reaches a predefined threshold. Given a target threat model under test, BASFuzz takes user input, consisting of a prompt $P$ and an original text example $T_{ori}$, as seed input. During the preprocessing stage, the input text undergoes a filtering operation to extract a set of perturbable words $\tilde I$ (cf. Section~\ref{sec4_3}), ensuring that subsequent perturbations are applied only to semantically sensitive words, thereby improving testing efficiency. In the perturbation stage (cf. Section~\ref{sec4_4}), BASFuzz queries the \textbf{open multilingual wordNet (OMW)} to retrieve synonyms, near-synonyms, and hypernyms\&hyponyms for each word in the perturbable set $\tilde I$. It then uses LLM-based word embeddings to compute semantic similarity for the retrieved candidates and replaces original words with semantically similar alternatives to generate variants. This process builds a diverse perturbation space for the source text, laying the foundation for high-quality and multilingual-adaptive test cases.

BASFuzz then enters the fuzzing loop (cf. Section~\ref{sec4_5}), performing an iterative search over the perturbation space using an innovative beam-annealing search. Beam search maintains multiple perturbation paths in parallel within the high-dimensional input space, increasing coverage of the input space. To avoid local optima, a simulated annealing-based probabilistic acceptance mechanism allows the retention of candidate test cases with temporarily lower scores, enhancing global exploration. Furthermore, BASFuzz computes the entropy of the perturbation distribution to dynamically adjust beam width, achieving an adaptive balance between perturbation diversity and computational resource allocation. An elitism mechanism is also introduced to probabilistically track top-performing candidates across iterations, improving testing efficiency. After multiple rounds of selection, perturbation, and convergence, BASFuzz outputs the optimal candidate as the final test case.

\subsection{Objective Function}\label{sec4_2}
In BASFuzz, we formulate robustness testing as a combinatorial optimization problem, where the objective is to maximize performance degradation of the model output with input perturbations as variables. Specifically, for evaluating the robustness of LLM-based NLP software on machine translation tasks, given an example dataset $\mathcal{D} = \{ (T_x^i,T_y^i)\} _{i = 1}^N$ and an original prompt $P$ such as ``Please translate the following German sentence into English:'', we apply perturbations $\delta$ to the combined input $[P; T_x]$ to generate adversarial test cases. To address Challenge \#1, under the constraint set $C$ and a predefined perturbation budget, BASFuzz searches for perturbation paths that cause the threat model output to deviate as much as possible from the reference translation $T_y$. The objective function can be formalized as:
\begin{equation}
\operatorname*{\textit{arg\,max}}_{\delta \in C}
    \mathbb{E}_{(T_x;T_y) \in \mathcal{D}}
    \mathcal{L}[\mathcal{F}([P;T_x] + \delta ),T_y]
\label{eq3}
\end{equation}
where $\mathcal{L}(\cdot, \cdot)$ is the loss function measuring the difference between the software-generated translation and the reference. In this study, we adopt the negative BLEU score as $\mathcal{L}$. BLEU, proposed by Papineni et al.~\cite{10.3115/1073083.1073135}, is a standard automatic evaluation metric for machine translation. It measures similarity between machine-generated and reference translations by computing $n$-gram overlap. The BLEU computation consists of two steps:

(1) $N$-gram precision:
For a given $n$-gram length, count the overlapping $n$-grams between the generated and reference translations, divided by the total number of $n$-grams in the generated translation. Formally expressed, for an $n$-gram set $G$, the precision is:
\begin{equation}
{P_n} = {{\sum\nolimits_{g \in G} {Coun{t_{clip}}(g)} } \over {\sum\nolimits_{g^{\prime} \in G^{\prime}} {Coun{t_{clip}}(g^{\prime})} }}
\label{eq4}
\end{equation}
where $Coun{t_{clip}}(g)$ is the clipped count of $g$ in the generated translation, limited by its maximum occurrence in the reference, and $G^{\prime}$ denotes all $n$-grams in the generated translation.

(2) Weighted geometric mean with \textbf{brevity penalty (BP)}:
Combine multiple $n$-gram precisions using a weighted geometric mean and apply BP to penalize overly short outputs:
\begin{equation}
BLEU = BP \cdot \exp (\sum\limits_{n = 1}^N {{w _n}\log {P_n}} )
\label{eq5}
\end{equation}
where $w _n$ are typically uniform weights. BP is computed as:
\begin{equation}
BP =
\begin{cases}
1, & \text{if } c > r \\
\exp(1 - r / c), & \text{if } c \leq r
\end{cases}
\label{eq6}
\end{equation}
with $c$ and $r$ denoting the lengths of the generated and reference translations, respectively. As a mainstream metric for machine translation, BLEU balances precision and fluency, scales well for large-scale robustness testing, and is widely adopted in translation benchmarks, ensuring comparability and acceptance. In the robustness evaluation of LLM-based NLP software, a higher BLEU score indicates that the output is closer to the reference in wording and structure. In comparison, a lower score reflects degraded translation quality. Therefore, BASFuzz defines $\mathcal{L}$ as the negative BLEU score and generates input variants that reduce BLEU below a given threshold to expose robustness flaws. This strategy is not limited to machine translation and can be extended to NLU tasks such as text classification by selecting appropriate task-specific evaluation metrics (cf. Section~\ref{sec6_5}).
\subsection{Filtering Criteria}\label{sec4_3}
In mutation-based fuzzing, filtering criteria are a critical first step to ensure both the validity and efficiency of generated variants. BASFuzz introduces stop-word filtering prior to constructing the perturbation space. This step removes stop words from the input text so that subsequent perturbations focus on core words with substantive semantic impact. Stop words are high-frequency functional words in natural language text that contribute little to the main semantics, such as ``the'', ``and'', ``to'' in English, or ``und'', ``der'', ``zu'' in German. They are often ignored in various NLP tasks because their role in semantic modeling and task discrimination is limited. Given an input sentence $I = [{w_1},{w_2}, \ldots ,{w_n}]$ and a predefined stop-word set $S$, the filtering operation can be defined as:
\begin{equation}
\tilde I = [{w_i}|{w_i} \in I \wedge {w_i} \notin {S_{stop}}]
\label{eq7}
\end{equation}
where $\tilde I$ denotes the retained word set containing only non-stop words. For example, for the sentence ``Please translate the following German sentence into English:'', the filtering step removes ``the'' and ``into'', retaining words such as ``Please'', ``translate'', and ``German'' for perturbation.

In BASFuzz, we adopt the standard stop-word lists provided by the NLTK library~\cite{10.3115/1118108.1118117}. Applying stop-word filtering before perturbation reduces the size of the perturbation space. For a sentence of length $n$ with an average stop-word rate $\rho$, the effective perturbation space is reduced to $(1-\rho)^n$ of the original, improving testing efficiency. Furthermore, perturbing stop words often degrades text fluency or introduces unnecessary noise; excluding them helps maintain the stealthiness of generated test cases.

\subsection{Perturbation Space Construction}\label{sec4_4}
Perturbation space construction is the foundation for generating high-quality test cases. The perturbation space defines the range of input variants BASFuzz can explore during automated testing, and its diversity and validity directly determine the effectiveness of robustness evaluation. BASFuzz employs a two-stage perturbation strategy: multilingual WordNet-based lexical retrieval (cf. Section~\ref{sec4_4_1}) and LLM-based semantic constraints (cf. Section~\ref{sec4_4_2}).
\subsubsection{Perturbation Words Retrieval}\label{sec4_4_1}
To build a diverse perturbation space, BASFuzz uses OMW as the core lexical resource. Unlike traditional WordNet resources limited to English, OMW integrates Princeton WordNet with semantic mappings across dozens of languages, providing a unified multilingual concept hierarchy. This enables the generation of semantically related lexical perturbations in multilingual tasks such as machine translation. Given an input text $I$, for each retained word $w_i$ after filtering, BASFuzz retrieves a candidate word set $S(w_i)$ from OMW. Three semantic perturbation modes are applied:

(1) Synonym Substitution: For word $w_i$, define a synonym set:
$
{S_{syn}}({w_i}) = \{ s \mid s \in Synonyms({w_i}),{\rm{ }}s \ne {w_i}\}
$.
All members are semantically equivalent to $w_i$. For example, ``translate'' may map to ``render'' or ``interpret.''

(2) Near-Synonym Substitution: Retrieve words with short semantic paths to $w_i$ (e.g., co-occurrence context or shared composite meaning), and define the set:
$
    S_{near}(w_i) = \{ s \mid s \in \text{nearSynonyms}(w_i) \} 
$.
For example, ``error'' may be perturbed into ``fault'' or ``issue'', allowing evaluation under subtle shifts in meaning.

(3) Hypernym/Hyponym Substitution: Based on OMW's concept hierarchy, retrieve direct hypernyms and hyponyms of $w_i$:
$
{S_{hyp}}({w_i}) = \{ s \mid s \in Hypernyms({w_i}) \cup Hyponyms({w_i})\}
$.
For example, ``vehicle'' may be replaced by ``car'' or ``transport'' to test generalization and reasoning ability of the threat model.

The final candidate perturbation set for each word is the union:
$S({w_i}) = {S_{syn}}({w_i}) \cup {S_{near}}({w_i}) \cup {S_{hyp}}({w_i})$.
Compared to most WordNet-based fuzzing approaches, BASFuzz benefits from OMW's finer-grained sense hierarchy and higher semantic resolution. This design accommodates the complex requirements of NLG robustness testing and mitigates perturbation sparsity and semantic drift caused by limited single-language resources. By integrating the three perturbation modes, BASFuzz not only enriches the perturbation space but also improves its applicability to robustness testing for multilingual NLP software.

\subsubsection{LLM-based Encoding}\label{sec4_4_2}
After retrieving candidate perturbation words, BASFuzz incorporates a word embedding LLM to evaluate semantic similarity, enhancing both the semantic precision of variants and their multilingual adaptability. Although OMW-based lexical retrieval provides a rich candidate set, its static dictionary-style semantic relations have inherent limitations for NLG tasks. From a linguistic perspective, OMW defines similarity primarily through hierarchical lexical relations and fails to capture context-sensitive semantic dynamics fully. In addition, the fixed symbolic network is prone to introducing inappropriate substitutions when handling polysemous words or phrase-level expressions subject to semantic drift. To address these issues, BASFuzz employs a Gemma2-based embedding model~\cite{DBLP:journals/corr/abs-2409-15700} to encode both the original word $w_i$ and all candidate words into high-dimensional vectors. Trained on large-scale multilingual corpora, this model has strong contextual modeling capability, and its output vectors implicitly capture the semantic distribution of words in diverse contexts. Given a word $w_i$ and its candidate set $S({w_i}) = \{ {s_1},{s_2}, \ldots ,{s_k}\} $, BASFuzz obtains their embeddings ${v_{{w_i}}} = E({w_i})$ and ${v_{{s_j}}} = E({s_j})$ for $\forall j = 1, \ldots ,k$. Semantic similarity between the original word and each candidate is measured using cosine similarity:
\begin{equation}
sim({w_i},{s_j}) = {{{v_{{w_i}}} \cdot {v_{{s_j}}}} \over {\left\| {{v_{{w_i}}}} \right\| \times \left\| {{v_{{s_j}}}} \right\|}}
    \label{eq8}
\end{equation}
Candidates are ranked by similarity scores, and $w_i$ is replaced with the top-$K$ closest candidates to generate variants. By introducing LLM-based semantic constraints on top of OMW retrieval, BASFuzz expands perturbation diversity while maintaining text quality through semantic consistency, providing a high-quality perturbation space for the subsequent fuzzing loop.
\subsection{Fuzzing Loop}\label{sec4_5}
\begin{algorithm}[t]
    \footnotesize
    \caption{Fuzzing Loop in BASFuzz}
    \label{algorithm1}
    \begin{algorithmic}[1]
	\REQUIRE $T_{ori}$: original text example, $P$: user prompt, $b$: beam width, $tem_{SA}$: simulated annealing temperature.
	\ENSURE $I_{adv}$: adversarial test case.
        \STATE $beam$, $I^*$$\leftarrow$Join($P$, $T_{ori}$);\rcmt{Initialize the fuzzing loop}
        \STATE $indexOrder$$\leftarrow$argsort(WIR($beam$), order=descend);
        \STATE $tem_{SA}$$\leftarrow$1;
        \STATE $iterNum$$\leftarrow$0;
        \WHILE{$iterNum$$<$|$indexOrder$|}
        \STATE candidateText$\leftarrow$Perturbation($beam$, $indexOrder$);\rcmt{Generate candidate variants}
        \FOR{each $I'$$\in$candidateText}
        \STATE $p_{SA}$$\leftarrow$Compute simulated annealing acceptance probability via $Eq.\ref{eq12}$;
        \IF{rand( )$<$$p_{SA}$}
        \STATE $tempBeam$$\leftarrow$$tempBeam$$\cup$$I'$;\rcmt{Accept with simulated annealing probability}
        \ENDIF
        \ENDFOR
        \STATE $iterNum$$\leftarrow$$iterNum$+1;
        \STATE Update $I^*$, $tem_{SA}$ via $Eq.\ref{eq13}$;
        \IF{$I_{adv}$ in $tempBeam$}
        \RETURN $I_{adv}$$\leftarrow$$I^*$\rcmt{Output successful test case}
        \ENDIF
        \STATE $H_{beam}$$\leftarrow$Entropy($\mathcal{L}$($tempBeam$));
        \STATE Update $b$ with $H_{beam}$ via $Eq.\ref{eq16}$;\rcmt{Entropy-based pruning}
        \STATE Compute elite retention probability $p_e$ via $Eq.\ref{eq18}$;
        \IF{rand( )$<$$p_e$}
        \STATE $nextBeam$$\leftarrow$$nextBeam$$\cup$$I^*$;\rcmt{Probabilistically retain elite variant}
        \STATE $tempBeam$$\leftarrow$$tempBeam$$\setminus$$\{I^*\}$;
        \STATE $b$$\leftarrow$$b$-1;
        \ENDIF
        \STATE $beam$$\leftarrow$$nextBeam$$\cup$SoftSample($tempBeam$,$b$,$\mathcal{L}$($tempBeam$));        
        \ENDWHILE
        \RETURN $I_{adv}$$\leftarrow$$I^*$
 \end{algorithmic}
\end{algorithm}
In high-dimensional natural language perturbation spaces, efficient robustness testing relies on automated search processes with global exploration capability and adaptability to dynamic feedback. To address the degradation of fuzzing capability in NLG tasks identified in Challenge \#2, BASFuzz designs a fuzzing loop driven by beam annealing search. The execution workflow is illustrated in Algorithm~\ref{algorithm1}. This algorithm maintains multiple perturbation paths in parallel through a beam mechanism, incorporates simulated annealing criteria to escape local optima, and integrates entropy-based beam-width adjustment with elitism to enable efficient automated robustness testing.

The fuzzing loop concatenates the original text example $T_{ori}$ and the user prompt $P$ into the initial input $I_{ori}$, which serves as both the starting node and the global best result $I^*$ (Line 1). BASFuzz then computes \textbf{word importance ranking (WIR)} to determine the perturbation priority index $\textit{indexOrder}$ for each word (Line 2). Given $I_{ori} = \{w_1, w_2, \ldots, w_i, \ldots, w_n\}$ and a masked version $I_{ori}^i = \{w_1, w_2, \ldots, [\text{UNK}], \ldots, w_n\}$ where the $i$-th word $w_i$ is replaced with an unknown token, the perturbation effect of $w_i$ is measured by the loss difference between $I_{ori}$ and $I_{ori}^i$. The word importance score is thus defined as:
\begin{equation}
WIR({w_i}) = {\rm{softmax}}(\mathcal{L}(I_{ori}^i,{T_y}) - \mathcal{L}({I_{ori}},{T_y})) \cdot (\mathcal{L}(I_{ori}^i,{T_y}) - \mathcal{L}({I_{ori}},{T_y}))
    \label{eq9}
\end{equation}
By ranking words based on the BLEU loss, the more they influence the output loss, the higher their importance score. BASFuzz prioritizes $\textit{indexOrder}$ on positions with the greatest impact on the generated output, improving search efficiency and reducing redundant mutations on low-sensitivity words. The fuzzing loop is initialized by setting the simulated annealing temperature to 1 and the iteration counter to 0 (Lines 3–4).

The fuzzing loop continues execution until a successful test case $I_{adv}$ is found or $\textit{indexOrder}$ is fully traversed (Lines 5–27). In each iteration, BASFuzz sequentially selects the next high-importance word from $\textit{indexOrder}$ for perturbation and generates a batch of new candidate texts $\textit{candidateText}$ (Line 6). For each candidate variant $I^\prime$, BASFuzz computes the acceptance probability $p_{SA}$ based on the simulated annealing criterion and uses this probability to decide whether to include the variant in the temporary beam for the current round (Lines 7–12). After annealing-based selection of all candidates, the iteration counter is incremented, and both $I^*$ and the simulated annealing temperature $tem_{SA}$ are updated (Lines 13–14). If a successful test case is found within the current beam, the loop terminates immediately (Lines 15–17). To enhance search diversity and control exploration breadth, BASFuzz computes the entropy $H_{beam}$ of the loss distribution in the temporary beam during each iteration and dynamically adjusts the beam width based on information entropy, adapting resource allocation to the dispersion of the distribution (Lines 18–19). Next, with probability $p_e$, BASFuzz carries $I^*$ over to the next beam to avoid losing high-value candidates during the loop (Lines 20–25). After updating the beam width, BASFuzz applies weighted sampling using the loss scores of remaining variants to fill the beam to capacity, forming the next-generation beam (Line 26). Finally, the global best variant is output as the test case $I_{\text{adv}}$ (Line 28). We now present the detailed procedures of the four main modules in the fuzzing loop.

\subsubsection{Beam Generation}
In the fuzzing loop of BASFuzz, beam generation is the key step that enables parallel exploration of the high-dimensional perturbation space. Compared with the single-path greedy strategy, beam search maintains multiple perturbation paths in parallel during each iteration, increasing perturbation diversity and effectively exposing potential robustness flaws in LLM-based NLP software. Given the current beam ${B^t} = \{ I_1^t,I_2^t, \ldots ,I_b^t\} $, where $b$ is the current beam width and each $I_j^t$ represents an input text in this iteration, BASFuzz selects one word position $w_i$ per round based on $\textit{indexOrder}$ obtained from WIR and generates a set of candidate variants:
\begin{equation}
{\rm{Replace}}(I_j^t,{w_i}) = \{ I_j^t[{w_i} \to c]|c \in Cand({w_i})\} 
    \label{eq10}
\end{equation}
where $Cand(w_i)$ is the perturbation word set constructed in the perturbation space. All generated variants are merged to form the next-step candidate text set:
\begin{equation}
{\rm{candidateText}} = \bigcup\limits_{I_j^t \in {B^t}} {{\rm{Replace}}(I_j^t,{w_i})}
    \label{eq11}
\end{equation}

For each candidate text generated through perturbation, subsequent steps combine simulated annealing probabilities with objective function scores to determine which variants advance to the next beam. For NLP software, the input space is discrete, sparse, and subject to a combinatorial explosion. Beam generation mitigates premature convergence to local optima by preserving and expanding multiple candidate paths at each iteration. This design ensures that robustness testing meets the fundamental requirement of testing effectiveness. The multi-path parallel expansion significantly increases the likelihood of reaching complex perturbation trajectories and enables deeper probing of the software's decision boundaries, thereby facilitating more comprehensive robustness evaluation.
\subsubsection{Selection via Simulated Annealing}
Although beam generation provides diversity for test cases, in NLG tasks, the nonlinear decision boundaries of LLMs and their high sensitivity to input perturbations make beam search guided by the greedy strategy prone to local optima. To address this, BASFuzz introduces a simulated annealing-based probabilistic acceptance strategy. During high-temperature phases, it accepts a certain degree of quality degradation in variants, enhancing global search capability and mitigating the early convergence to local optima inherent in traditional greedy search. For each candidate text $I^\prime$, BASFuzz computes its loss increment relative to the original input using the objective function:
$\Delta \mathcal{L} = \mathcal{L}(I^\prime,{T_y}) - \mathcal{L}({I_{ori}},{T_y})$.
A temperature parameter $\textit{tem}_{SA}$ is then introduced, and the acceptance of a candidate is determined using the following probability function:
\begin{equation}
    p_{SA}(I^\prime) = 
    \begin{cases} 
        1, & \Delta \mathcal{L} > 0 \\ 
        \exp\left(\frac{\Delta \mathcal{L}}{\textit{tem}_{SA}}\right), & \Delta \mathcal{L} \le 0 
    \end{cases}
    \label{eq12}
\end{equation}
That is, when the candidate $I^\prime$ achieves a better objective score, it is always accepted; otherwise, it is accepted with probability $\exp\left(\frac{\Delta \mathcal{L}}{\textit{tem}_{SA}}\right)$, allowing some degraded candidates to enter subsequent search rounds. BASFuzz applies a logarithmic decay schedule to gradually reduce the annealing temperature, guiding the search from exploration to convergence:
\begin{equation}
te{m_{SA}}(t) = {{te{m_{SA}}(0)} \over {1 + \gamma  \cdot \ln (1 + t)}}
    \label{eq13}
\end{equation}
where $\gamma$ is the cooling factor and $te{m_{SA}}(0)$ is the initial temperature. In the early high-temperature phase, BASFuzz relaxes the acceptance threshold for suboptimal perturbations, encouraging paths that cross larger semantic gaps to escape local optima. As iterations progress and the temperature decreases, the search converges and increasingly favors better perturbations. This balance between early-stage exploration and late-stage fine convergence makes the approach particularly suitable for black-box robustness testing scenarios with no gradients and no interpretable internal states.

\subsubsection{Entropy-based Pruning}
In standard beam search, the beam width $b$ is a fixed hyperparameter that remains constant throughout the search. Formally, at each iteration, only the top-$b$ candidates are retained based on their scores. While this approach controls search complexity to some extent, it exhibits significant limitations for NLG tasks. In the early stages of the fuzzing loop, the perturbation space is not yet sufficiently explored; a fixed small beam width restricts variant diversity and increases the risk of local optima. In later stages, when candidates begin to converge, a fixed large beam width wastes resources on redundant perturbations. More critically, during different phases of testing, the BLEU score distribution of candidate texts changes dynamically. A single static beam width cannot adapt to the structural complexity of the perturbation space, ultimately reducing the efficiency of robustness testing.

To address the shortcomings of a fixed beam width, BASFuzz draws inspiration from information theory and proposes an entropy-based beam pruning strategy. The key idea is that the entropy of the score distribution of candidate texts can measure the diversity and uncertainty of the current perturbation space. When the score distribution entropy is high, large mutations exist among different paths, indicating that the beam width should be expanded to strengthen global exploration. When the entropy is low, candidate paths become homogeneous, allowing the beam width to be reduced, which in turn improves fuzzing efficiency and convergence speed. Given the current beam ${B^t} = \{ I_1^t,I_2^t, \ldots ,I_b^t\} $, BASFuzz first normalizes the loss scores of all candidate texts to obtain a probability distribution:
\begin{equation}
{p_j} = {{\mathcal{L}(I_j^t,{T_y})} \over {\sum\limits_{i = 1}^b {\mathcal{L}(I_i^t,{T_y})} }}
    \label{eq14}
\end{equation}
then computes the information entropy of the current beam using Shannon's formula:
\begin{equation}
H_{beam} = - \sum_{i=1}^{b} p_i \log (p_i + \epsilon)
    \label{eq15}
\end{equation}
where $\epsilon$ is a small positive constant to avoid the logarithm of zero, a higher $H_{beam}$ indicates greater diversity among candidate texts and higher exploration potential. Unlike traditional methods with fixed beam width, BASFuzz adjusts $b$ dynamically based on $H_{beam}$:
\begin{equation}
{b_{t + 1}} = \max ({b_{\min }},\min ({b_{\max }},{b_t} \cdot (1 + {{{H_{beam}}} \over {{b_{\max }}}}),{b_t} + \sigma ))
    \label{eq16}
\end{equation}
where $b_{\min}$ and $b_{\max}$ are the minimum and maximum beam widths to keep fuzzing overhead within a controllable range; ${b_t} \cdot (1 + {{{H_{beam}}} \over {{b_{\max }}}})$ is a scaling factor to expand the beam width in high-entropy phases; and $\sigma$ is a smoothing term to avoid overly rapid contraction in low-entropy phases. Compared with standard beam search, the entropy-based pruning strategy overcomes the inherent limitation of static beam width. It integrates the information-theoretic principle into NLP software robustness testing, improving both the effectiveness and efficiency of large-scale fuzzing.

\subsubsection{Elitism}
After entropy-based beam width adjustment, BASFuzz can dynamically allocate search resources according to the uncertainty of the variant distribution. However, the responses of LLM-based NLP software often exhibit nonlinear jumps. Dynamic beam width alone is difficult to mitigate against accidentally discarding promising perturbation sequences during simulated annealing selection, especially near decision boundaries. To address this issue, we introduce an elitism mechanism to track the best perturbation path discovered so far robustly. Conventional elitism unconditionally carries the current global best candidate into the next fuzzing iteration. While this guarantees that the optimal perturbation path is not lost, it can cause all variants to converge prematurely to a local extremum during the early stages of testing, reducing the ability to explore the broader perturbation space. This behavior is particularly detrimental in software testing, as it risks missing robustness flaws hidden in alternative subspaces of perturbations.

To mitigate this over-convergence, BASFuzz employs probabilistic elitism to balance persistent tracking of the global best with the exploration of diverse perturbations. For the current global best input $I^*$ generated in the fuzzing loop, its soft retention probability relative to all candidates is defined as:
\begin{equation}
{p^*} = {{\exp (\mathcal{L}({I^*},{T_y}))} \over {\sum\limits_{i = 1}^b {\exp (\mathcal{L}(I_i^t,{T_y}))} }}
    \label{eq17}
\end{equation}
The elitism retention probability is then computed using a base elitism rate $p_0^*$ and the soft retention probability $p^*$:
\begin{equation}
{p_e} = p_0^* + (1 - p_0^*) \cdot {p^*}
    \label{eq18}
\end{equation}
when the global best is significantly superior to other candidates, $p^*$ is high, which further increases $p_e$ and reinforces focus on the optimal perturbation path. When $I^*$ is only marginally better, $p_e$ decreases accordingly, allowing BASFuzz to retain $I^*$ with high but not absolute probability, encouraging diverse perturbation paths to enter the next fuzzing iteration. By establishing an elastic transition between local exploitation and global exploration, the probabilistic elitism mechanism ensures the global best is not easily lost while preventing early convergence to a single path. This mechanism guarantees that even when simulated annealing accepts a large number of suboptimal perturbations, the current best sequence is preserved in a steady state, enhancing the fuzzing capability of BASFuzz.

After elitism retention, BASFuzz performs soft-probability weighted sampling, similar to $p^*$, to fill the remaining beam width without replacement. The fuzzing loop continues until termination, and the final global best $I^*$ is output as the test case generated by BASFuzz.

\section{Experimental Settings}\label{sec5}
To evaluate the effectiveness, efficiency, and practical value of BASFuzz in robustness testing for LLM-based NLP software, we design and conduct extensive experiments. All code and data used in the experiments are released in a reproducible repository~\cite{BASFuzz} to facilitate community replication and further research. The experimental design is structured around the following five research questions (RQs):
\begin{itemize}
\item \textit{RQ1:} How effective is BASFuzz in generating high-quality test cases?
\item \textit{RQ2:} How lightweight is BASFuzz when performing robustness testing?
\item \textit{RQ3:} What is the contribution of different components to BASFuzz's testing effectiveness?
\item \textit{RQ4:} How transferable are BASFuzz-generated test cases across different threat models?
\item \textit{RQ5:} Can BASFuzz maintain robustness testing effectiveness when applied to the NLU task?
\end{itemize}

\subsection{Datasets}
For the first four RQs, we select machine translation as the NLG task scenario. As a representative task in the NLG domain, machine translation produces open-ended text sequences without fixed-category constraints, fully reflecting the semantic flexibility and diversity challenges faced by LLM-based software in generative tasks. In addition, machine translation is highly sensitive to subtle input perturbations, making it well-suited for validating robustness testing methods. Specifically, we adopt the WMT16 dataset\footnote{\url{https://huggingface.co/datasets/wmt/wmt16}}~\cite{bojar-EtAl:2016:WMT1}, a widely recognized benchmark in the machine translation field. WMT16 has long been used in international machine translation competitions and contains high-quality parallel corpora across multiple language pairs. It features strong linguistic diversity, rich syntactic structures, and a broad topical distribution, making it a representative choice for robustness evaluation in NLG tasks. From WMT16, we select three language pairs as the source datasets: Czech-to-English (CS2EN), German-to-English (DE2EN), and Russian-to-English (RU2EN). Czech, as a Slavic language, exhibits rich morphological mutation; German has complex grammatical structures and long compound words; and Russian differs from English in word order. These characteristics allow us to validate BASFuzz's effectiveness under multilingual settings.

For RQ5, we extend the study to NLU tasks and use text classification as a typical representative. Text classification requires the software to perform semantic understanding of input examples and accurately assign them to predefined category labels, exemplifying structured prediction in LLM-based software. This setup facilitates the investigation of BASFuzz's scalability across different task types. We employ several widely used datasets in text classification: Financial Phrasebank (FP)\footnote{\url{https://huggingface.co/datasets/financial_phrasebank}}~\cite{malo2014good}, AG's News\footnote{\url{https://s3.amazonaws.com/fast-ai-nlp/ag\_news\_csv.tgz}}~\cite{NIPS2015_250cf8b5}, and MR\footnote{\url{https://huggingface.co/datasets/rotten_tomatoes}}~\cite{pang2005seeing}. Financial Phrasebank contains short sentences with sentiment labels extracted from 10,000 real-world financial news articles and analysis reports. It is widely used in financial sentiment classification research and reflects the impact of highly structured domain-specific language on the scalability of testing methods. AG's News consists of news headlines and summaries across topics such as world news, sports, business, and technology. It includes 496,835 articles from over 2,000 news sources and is used to examine BASFuzz's capability in multi-class classification tasks and general-topic scenarios. MR is a dataset focused on sentiment polarity detection in movie reviews. It contains 10,662 review snippets written by professional film critics from Rotten Tomatoes, representing a fine-grained and challenging environment for robustness testing.

\subsection{Threat models}
In current software engineering practice, most LLM-based NLP software tends to perform inference by directly invoking APIs of pre-trained LLMs rather than training task-specific models or building complex post-processing pipelines. This approach reduces deployment and maintenance costs and enables rapid integration of advanced natural language capabilities into various products. However, it also implies that the robustness of the software system is largely constrained by the underlying LLM on which it depends. Therefore, treating the LLM as the threat model in robustness testing helps reveal fundamental vulnerabilities and provides a more intrinsic assessment of the robustness of higher-level software applications. To evaluate the effectiveness of BASFuzz in diverse LLM-based NLP environments, we select five open-source LLMs as threat models. This choice also addresses the stringent requirements for auditability and compliance of software components in high-security domains such as finance and healthcare, providing greater transparency and regulatory assurance. Compared with closed-source LLMs, the use of open-source models enhances the reproducibility of experimental results and provides a solid foundation for verifiability in the research process. We select five popular LLMs, all accessible through the Hugging Face platform: Mistral-7B-Instruct-v0.3\footnote{\url{https://huggingface.co/mistralai/Mistral-7B-Instruct-v0.3}}~\cite{jiang2023mistral}, Phi-4\footnote{\url{https://huggingface.co/microsoft/phi-4}}~\cite{DBLP:journals/corr/abs-2412-08905}, InternLM2.5-20b-chat\footnote{\url{https://huggingface.co/internlm/internlm2_5-20b-chat}}~\cite{DBLP:journals/corr/abs-2403-17297}, Yi-1.5-34B-Chat\footnote{\url{https://huggingface.co/01-ai/Yi-1.5-34B-Chat}}~\cite{DBLP:journals/corr/abs-2403-04652}, and Llama-3-70B-Instruct\footnote{\url{https://huggingface.co/meta-llama/Meta-Llama-3-70B-Instruct}}~\cite{llama3modelcard}. These models span small, medium, and ultra-large parameter scales and cover current mainstream LLM architectures, enabling the evaluation of BASFuzz under varying levels of linguistic reasoning complexity.

\subsection{The Settings of Test Generation}
All experiments in this study are conducted on an Ubuntu 22.04.1 LTS operating system. The hardware configuration consists of two Intel Xeon Platinum 8358 processors with 32 physical cores running at 2.60 GHz, NVIDIA A100 Tensor Core GPUs, and 1 TB of physical memory. All baselines and threat models are deployed and configured following their official documentation. Each experiment is independently repeated six times, and the arithmetic mean of all evaluation metrics is reported to ensure statistical robustness of the conclusions. 
For every experiment, the threat model randomly selects 1,000 text examples from the dataset as user inputs to guarantee diversity of input types and enhance the representativeness and reliability of results.

Following common practice in machine translation security testing~\cite{10.1145/3377811.3380420}, we set the BLEU threshold in the objective function to 0.2. A generated input variant is considered a successful test case only when it causes the software output's BLEU score to drop below 0.2. BLEU is computed using the widely adopted 4‑gram configuration~\cite{10.3115/1073083.1073135}, simultaneously evaluating 1‑gram to 4‑gram segments to balance quality assessment for both short and long text generation. During perturbation space construction, we set the number of candidate words $K$ = 10 for each word selected for perturbation to ensure sufficient diversity while preventing exponential growth of the perturbation space. In the fuzzing loop stage, all hyperparameters are tuned through pilot experiments and informed by industry best practices. Table~\ref{tab1} summarizes the main hyperparameter settings.
\begin{table}[th]
\caption{Hyperparameter settings in the fuzzing loop.}
\label{tab1}
\begin{tabular}{c|c|c}
\hline
Hyperparameter           & Symbol     & Value             \\ \hline
Cooling factor           & $\gamma$   & 0.3               \\
Entropy smoothing        & $\epsilon$ & $1\times10^{-10}$ \\
Initial beam width       & $b_{0}$    & 2                 \\
Min beam width           & $b_{min}$  & 2                 \\
Max beam width           & $b_{max}$  & 6                 \\
Beam width increment     & $\sigma$   & 1                 \\
Elitism base probability & $p_0^*$     & 0.9               \\ \hline
\end{tabular}
\end{table}

\subsection{Baselines}
To evaluate the performance of BASFuzz in testing the robustness of LLM-based NLP software, we select six recently proposed or widely adopted testing methods as baselines. Since most existing automated testing approaches are designed for DNN-based NLP software rather than directly targeting black-box LLM-based software, relevant baselines in this setting are limited. To bridge this gap, we adapt several methods originally developed for DNN-based NLP software by modifying their interaction mechanisms and objective functions for application to the emerging scenario studied in this paper. In preliminary experiments, we observe that population-based metaheuristic search algorithms, such as genetic algorithm~\cite{pmlr-v161-wang21a} and particle swarm optimization~\cite{10298415}, incur extremely high testing costs, with the average generation time per successful test case exceeding 500 minutes. Although these methods have demonstrated strong global search capabilities in small-scale DNN-based software testing, their application to large-scale LLM-based robustness testing violates the fundamental goal of automated testing, which is to improve development efficiency and maintainability. Therefore, we introduce an efficiency threshold in baseline selection, including only methods whose average generation time per successful test case does not exceed 180 minutes. 
For the first four RQs, we compare BASFuzz against six representative methods: ABS~\cite{xiao2024assessing}, ABFS~\cite{xiao2025abfs}, GreedyFuzz, VFA~\cite{10.24963/ijcai.2024/730}, Seq2Sick~\cite{Cheng_Yi_Chen_Zhang_Hsieh_2020}, and MORPHEUS~\cite{tan-etal-2020-morphin}. Among them, GreedyFuzz is a comparative method designed in this study based on standard fuzz testing and employing a classical greedy search strategy.
For the scalability study in RQ5, we note that VFA, Seq2Sick, and MORPHEUS are primarily designed for machine translation and are not directly applicable to NLU tasks. Therefore, we include two widely used testing methods, TextFooler~\cite{Jin_Jin_Zhou_Szolovits_2020} and PWWS~\cite{ren-etal-2019-generating}, as supplementary baselines. TextFooler generates test cases using a deletion-based selection mechanism that targets words with the greatest impact on the final decision while aiming to preserve semantic similarity. PWWS extends synonym substitution with a new word replacement order determined by word importance and classification probability, also employing a greedy search strategy.

\subsection{Evaluation Metrics}
In the experiments, we use six metrics to evaluate the testing effectiveness of BASFuzz, the quality of generated test cases, and the testing efficiency of BASFuzz. Following prior work~\cite{10.1145/3597503.3639121}, we perform the Mann-Whitney U-test~\cite{497e1044-d5b0-30a9-b230-3ca0f10d6f6c} to calculate the statistical significance of the experimental results for RQ1 and RQ5, with the significance level set to $\alpha$ = 0.05.

(1) \textit{Success Rate (S-rate)}~\cite{morris2020textattack2} measures the proportion of test cases generated by a testing method that successfully mislead the threat model's output over the total number of tested examples. In this study, it is defined as:
\begin{equation}
\text{S-rate}=\frac{N_{suc}}{N}\times100\%
\end{equation}
where $N_{suc}$ is the number of test cases that successfully mislead the threat model, and $N$ is the total number of input texts for the current testing method. A higher S-rate indicates greater effectiveness of the testing method in evaluating software robustness.

(2) \textit{Change Rate (C-rate)}~\cite{morris2020textattack2} measures the average proportion of perturbed words relative to the original input text. It is defined as:
\begin{equation}
\text {C-rate }=\frac{1}{N_{suc}} \sum_{k=1}^{N_{suc}} \frac{\operatorname{diff} I_k}{\operatorname{len}\left(I_k\right)}\times100\%
\end{equation}
where $\text{diff}(I_k)$ denotes the number of replaced words in input text $I_k$, and $\text{len}(\cdot)$ denotes the sequence length. A higher C-rate implies stronger perturbations, which may reveal more robustness flaws but can reduce the stealthiness of test cases.

(3) \textit{Perplexity (PPL)}~\cite{morris2020textattack2} is used to evaluate the fluency of generated test cases. PPL is defined as the exponential of the average negative log-likelihood of the input text. For a tokenized input $I = (w_1, w_2, \ldots, w_n)$, it is computed as:
\begin{equation}
\operatorname{PPL}(I)=\exp \left\{-\frac{1}{n} \sum_i^n \log p_\theta\left(w_i \mid w_{<i}\right)\right\}
\end{equation}
where $\log p_\theta(w_i \mid w_{<i})$ is the log-likelihood of token $w_i$ given the preceding tokens under a language model. Intuitively, the more fluent a test case is for the language model used to compute PPL, the less likely it is to introduce confusion.

(4) \textit{Grammar Errors (G-E)}~\cite{morris2020textattack2} represent the average number of grammatical errors per successful test case generated by the testing method. It is defined as:
\begin{equation}
\text {G-E}= \frac{1}{{N_{suc}}}\sum\limits_{k = 1}^{{N_{suc}}} {{E_k}} 
\end{equation}
where $E_k$ is the number of grammatical errors in the $k$-th test case. We use the open-source grammar and spell-checking tool LanguageTool to compute $E_k$. Excessive grammatical errors can degrade the text quality of test cases and reduce the accuracy and validity of testing results.

(5) \textit{Time Overhead (T-O)}~\cite{morris2020textattack2} measures the average time required by the testing method to generate one successful test case.

(6) \textit{Query Number (Q-N)}~\cite{morris2020textattack2} measures the average number of queries to the threat model needed to generate one successful test case. Together, Q-N and T-O reflect the overall efficiency of the testing method.

\section{Results and Analysis}\label{sec6}
\subsection{RQ1: How effective is BASFuzz in generating high-quality test cases?}
We use the success rate to evaluate the effectiveness of test cases generated by BASFuzz and analyze perturbation stealthiness and language fluency using change rate, perplexity, and grammar errors. Table~\ref{tab2} reports the comparative results across different datasets and threat models. In terms of testing effectiveness, MORPHEUS, ABS, and ABFS achieve average success rates of 72.363\%, 62.912\%, and 58.660\%, respectively, showing the strongest performance among current baselines. The best-performing baseline, MORPHEUS, reaches a success rate of 82.691\% on traditional DNN-based NLP software, indicating that existing methods retain some adaptability to LLM-based NLP testing scenarios but exhibit degraded fuzzing capability. BASFuzz achieves an average success rate of 77.701\% and consistently outperforms all baselines across every dataset and threat model. Taking the challenging cross-linguistic translation task RU2EN as an example, the six baselines achieve success rates of 55.853\%, 42.418\%, 2.448\%, 59.936\%, 77.446\%, and 38.279\% on InternLM2.5\_20B, while BASFuzz reaches 85.051\%, demonstrating further improvement in testing effectiveness. A higher success rate indicates that BASFuzz can cover a broader range of potential failure cases, providing more test cases for robustness evaluation and continuous improvement during deployment. Although MORPHEUS and ABS achieve success rates close to BASFuzz in some scenarios, they often sacrifice perturbation stealthiness and linguistic fluency of the generated test cases. Unlike traditional software, the decision-making process of LLM-based NLP software relies heavily on model knowledge rather than developer-defined system rules. Comprehensive robustness testing is therefore essential for establishing reliable software quality evaluation. The additional test cases discovered by BASFuzz identify critical risk regions in real-world applications, making the approach particularly suited for software safety evaluation in high-stakes domains.
\begin{table}[t]
\caption{Comparison of the quality of test cases generated by seven testing methods. We use \textbf{boldface} to indicate the best result under each specific setting and gray shading to highlight the performance of our method. An asterisk * denotes statistical significance (Mann-Whitney U-test, sig. level < 0.05).
}
\label{tab2}
\setlength{\tabcolsep}{0.7pt} 
\footnotesize
\begin{tabular}{c|c|ccc|ccc|ccc|ccc|ccc}
\hline
 &  & \multicolumn{3}{c|}{Mistral0.3\_7B} & \multicolumn{3}{c|}{Phi4\_14B} & \multicolumn{3}{c|}{InternLM2.5\_20B} & \multicolumn{3}{c|}{Yi1.5\_34B} & \multicolumn{3}{c}{Llama3\_70B} \\ \cline{3-17} 
\multirow{-2}{*}{Dataset} & \multirow{-2}{*}{Baseline} & S-rate & C-rate & PPL & S-rate & C-rate & PPL & S-rate & C-rate & PPL & S-rate & C-rate & PPL & S-rate & C-rate & PPL \\ \hline
 & ABS & 62.194 & 4.681 & 189.740 & 41.949 & 8.961 & 202.251 & 69.763 & 9.117 & 232.283 & 54.876 & 8.441 & 238.425 & 71.037 & 5.289 & 185.698 \\
 & ABFS & 59.344 & \textbf{3.042} & 183.233 & 38.444 & 7.894 & 193.607 & 66.189 & 6.932 & 180.284 & 52.254 & 7.666 & 205.078 & 73.988 & 4.716 & 183.453 \\
 & GreedyFuzz & 5.637 & 5.674 & 180.257 & 3.952 & 4.862 & 230.411 & 7.475 & 4.946 & 214.967 & 3.661 & 5.826 & 175.165 & 4.548 & 5.165 & 192.989 \\
 & VFA & 48.229 & 4.542 & 187.565 & 55.073 & 5.107 & 180.682 & 51.775 & 5.155 & 194.607 & 53.788 & 4.931 & 180.172 & 62.603 & 5.537 & 186.298 \\
 & MORPHEUS & 65.893 & 15.435 & 281.861 & 54.886 & 22.958 & 349.175 & 87.106 & 14.331 & 245.482 & 73.857 & 15.693 & 278.879 & 76.322 & 6.803 & 200.375 \\
 & Seq2Sick & 55.268 & 4.588 & 191.355 & 26.234 & 5.006 & 193.463 & 46.153 & 4.677 & 192.597 & 35.712 & 5.365 & 195.881 & 73.96 & 5.451 & 194.183 \\
\multirow{-7}{*}{CS2EN} & BASFuzz & \cellcolor[HTML]{C0C0C0}\textbf{68.551*} & \cellcolor[HTML]{C0C0C0}4.511 & \cellcolor[HTML]{C0C0C0}\textbf{173.828*} & \cellcolor[HTML]{C0C0C0}\textbf{64.601*} & \cellcolor[HTML]{C0C0C0}\textbf{3.553*} & \cellcolor[HTML]{C0C0C0}\textbf{170.092*} & \cellcolor[HTML]{C0C0C0}\textbf{88.611*} & \cellcolor[HTML]{C0C0C0}\textbf{3.712*} & \cellcolor[HTML]{C0C0C0}\textbf{136.955*} & \cellcolor[HTML]{C0C0C0}\textbf{75.432*} & \cellcolor[HTML]{C0C0C0}\textbf{3.739*} & \cellcolor[HTML]{C0C0C0}\textbf{166.643*} & \cellcolor[HTML]{C0C0C0}\textbf{79.497*} & \cellcolor[HTML]{C0C0C0}\textbf{3.324*} & \cellcolor[HTML]{C0C0C0}\textbf{181.264*} \\ \hline
 & ABS & 73.241 & 4.629 & 125.279 & 46.491 & 10.414 & 188.578 & 75.253 & 11.841 & 180.776 & 61.181 & 11.557 & 188.434 & 84.543 & 6.425 & 136.524 \\
 & ABFS & 69.542 & 4.415 & 121.433 & 37.439 & 8.999 & 166.078 & 72.616 & 7.392 & 163.389 & 54.776 & 10.617 & 125.489 & 81.083 & 4.982 & 121.389 \\
 & GreedyFuzz & 3.305 & 5.206 & 124.412 & 2.417 & 5.503 & 126.794 & 5.978 & 5.058 & 125.741 & 2.292 & 5.138 & 168.612 & 7.166 & 5.331 & 123.936 \\
 & VFA & 51.736 & 3.905 & 133.074 & 65.179 & 5.124 & 129.028 & 63.022 & 4.614 & 129.965 & 64.684 & 5.499 & 121.526 & 72.396 & 5.045 & 137.398 \\
 & MORPHEUS & 74.187 & 14.525 & 167.870 & 62.297 & 36.003 & 488.856 & 70.726 & 15.439 & 169.613 & 72.071 & 15.307 & 169.298 & 86.789 & 12.606 & 151.181 \\
 & Seq2Sick & 64.108 & 4.283 & 116.181 & 38.347 & 6.265 & 126.475 & 52.408 & 6.104 & 118.206 & 41.872 & 7.291 & 119.889 & 47.574 & 5.669 & 113.286 \\
\multirow{-7}{*}{DE2EN} & BASFuzz & \cellcolor[HTML]{C0C0C0}\textbf{77.368*} & \cellcolor[HTML]{C0C0C0}\textbf{3.118*} & \cellcolor[HTML]{C0C0C0}\textbf{115.104*} & \cellcolor[HTML]{C0C0C0}\textbf{72.651*} & \cellcolor[HTML]{C0C0C0}\textbf{4.743*} & \cellcolor[HTML]{C0C0C0}\textbf{124.292*} & \cellcolor[HTML]{C0C0C0}\textbf{87.796*} & \cellcolor[HTML]{C0C0C0}\textbf{4.291*} & \cellcolor[HTML]{C0C0C0}\textbf{116.331*} & \cellcolor[HTML]{C0C0C0}\textbf{80.167*} & \cellcolor[HTML]{C0C0C0}\textbf{4.385*} & \cellcolor[HTML]{C0C0C0}\textbf{116.664*} & \cellcolor[HTML]{C0C0C0}\textbf{90.335*} & \cellcolor[HTML]{C0C0C0}\textbf{3.198*} & \cellcolor[HTML]{C0C0C0}\textbf{108.883*} \\ \hline
 & ABS & 64.585 & 6.335 & 66.896 & 60.599 & 8.589 & 66.332 & 55.853 & 11.745 & 71.349 & 51.401 & 10.527 & 68.021 & 70.713 & 8.812 & 70.296 \\
 & ABFS & 61.917 & 5.456 & 66.545 & 61.042 & 3.563 & 64.423 & 42.418 & 13.491 & 68.803 & 49.358 & 6.544 & 64.824 & 59.497 & 7.869 & 64.768 \\
 & GreedyFuzz & 7.127 & \textbf{3.104} & 67.809 & 3.913 & 6.594 & 68.432 & 2.448 & 4.459 & 69.916 & 4.984 & 4.623 & 66.307 & 8.581 & 4.859 & 68.776 \\
 & VFA & 48.874 & 4.367 & 73.714 & 60.971 & 4.835 & 72.728 & 59.936 & 3.545 & 73.858 & 61.068 & 4.289 & 73.534 & 57.958 & 4.413 & 78.645 \\
 & MORPHEUS & 70.513 & 16.046 & 69.525 & 67.684 & 25.613 & 85.233 & 77.446 & 15.868 & 70.878 & 59.523 & 18.459 & 74.826 & 86.143 & 15.232 & 70.137 \\
 & Seq2Sick & 51.363 & 4.798 & 65.175 & 66.008 & 3.469 & 65.302 & 38.279 & 5.259 & 65.854 & 51.149 & 3.643 & 65.317 & 55.459 & 4.298 & 64.793 \\
\multirow{-7}{*}{RU2EN} & BASFuzz & \cellcolor[HTML]{C0C0C0}\textbf{70.525} & \cellcolor[HTML]{C0C0C0}4.216 & \cellcolor[HTML]{C0C0C0}\textbf{64.762*} & \cellcolor[HTML]{C0C0C0}\textbf{69.899*} & \cellcolor[HTML]{C0C0C0}\textbf{3.147*} & \cellcolor[HTML]{C0C0C0}\textbf{62.374*} & \cellcolor[HTML]{C0C0C0}\textbf{85.051*} & \cellcolor[HTML]{C0C0C0}\textbf{3.254*} & \cellcolor[HTML]{C0C0C0}\textbf{63.674*} & \cellcolor[HTML]{C0C0C0}\textbf{65.899*} & \cellcolor[HTML]{C0C0C0}\textbf{3.463*} & \cellcolor[HTML]{C0C0C0}\textbf{64.556} & \cellcolor[HTML]{C0C0C0}\textbf{89.138*} & \cellcolor[HTML]{C0C0C0}\textbf{3.572*} & \cellcolor[HTML]{C0C0C0}\textbf{63.635*} \\ \hline
\end{tabular}
\end{table}

The change rate reflects the stealthiness of perturbations introduced during testing and provides a more reproducible metric compared to subjective human evaluation. Owing to the mutation strategy combining a multilingual word network with LLM-based word embeddings, BASFuzz achieves consistently lower change rates in most cases. For example, although MORPHEUS shows strong performance in terms of success rate, its average change rate reaches 17.355\%, far exceeding BASFuzz's 3.748\%. Excessive perturbation often causes test cases to deviate from real-world inputs, reducing the practical significance of testing results. While ABFS and GreedyFuzz occasionally approach or slightly outperform BASFuzz in change rate on smaller models such as Mistral0.3\_7B, this is typically associated with the higher tendency of small-parameter LLMs to hallucinate. When a test case triggers hallucination, the LLM generates incorrect or irrelevant outputs, which can lead to BLEU scores that appear closer to the reference, thus reducing the apparent change rate. This phenomenon does not indicate better perturbation stealthiness but instead exposes robustness flaws in smaller LLMs that struggle to maintain stable semantic expression. For PPL as a quantitative measure of text fluency, BASFuzz also performs strongly, achieving an average perplexity of 115.271, lower than all compared baselines. This indicates that the generated test cases are more natural and deceptive. Higher linguistic fluency directly increases the threat to deployed software, as malicious users can induce software errors using inputs that are nearly indistinguishable from normal ones, thereby uncovering hidden security blind spots in real-world environments.

\begin{figure}[t]
    \centering
    \includegraphics[width=1\textwidth]{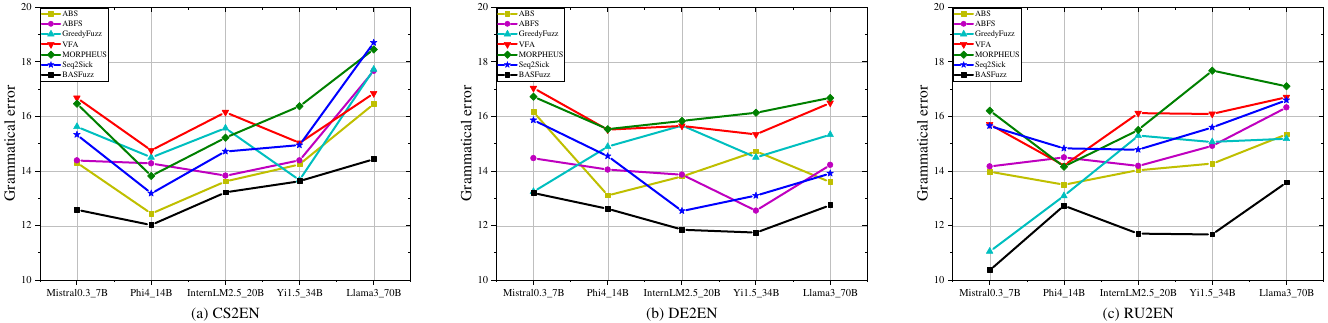}\\
    \caption{Comparison of the average number of grammatical errors per test case (want $\downarrow$).} 
    \label{Fig3}
\end{figure}

In NLP software testing, perturbations that preserve semantic integrity but degrade grammatical correctness can lead to false-positive test cases. To further evaluate test case quality and potential false alarms, we measure the number of grammatical errors in successful test cases. Figure~\ref{Fig3} shows the comparative results of seven testing methods. Although all methods generate test cases with some grammatical errors, moderate errors do not reduce the effectiveness of robustness testing. They can even help expose software robustness against non-standard inputs. BASFuzz consistently produces test cases with the fewest grammatical errors, averaging 12.541 errors per case, compared to 15.892 for VFA, which also leverages LLM-based semantic understanding. This demonstrates that BASFuzz can maintain testing effectiveness while minimizing false positives caused by minor linguistic defects, focusing evaluation precisely on the true robustness of the software.
\vspace{0.2cm}
\begin{mdframed}[backgroundcolor=gray!20, linecolor=black]
\textbf{Answer to RQ1:} 
BASFuzz significantly outperforms all baselines in success rate and generates higher-quality test cases. This indicates that BASFuzz is more effective at exposing robustness flaws, supporting more comprehensive and stealthy evaluations prior to software deployment. 
\end{mdframed}

\subsection{RQ2: How lightweight is BASFuzz when performing robustness testing?}
\begin{figure}[t]
    \centering
    \includegraphics[width=1\textwidth, height=0.14\textheight]{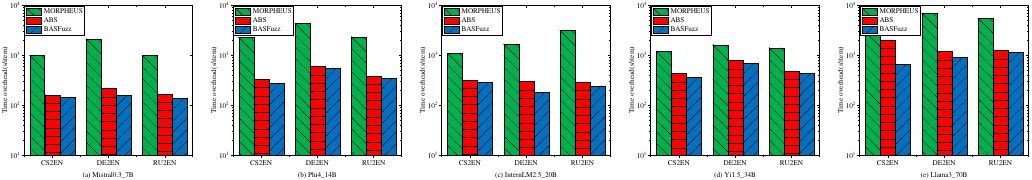}\\
    \caption{Results of test time overhead on different datasets and threat models (want $\downarrow$). }
    \label{Fig4}
\end{figure}

While ensuring the quality of test case generation, automated software testing must also focus on the execution efficiency of the testing process. Excessive time overhead and inference call costs can undermine the usability of testing methods in large-scale quality assurance processes. Therefore, this study evaluates the lightweight characteristics of BASFuzz in robustness testing, considering both testing time overhead and query number. To ensure the representativeness of the comparison, we select the two baseline methods, MORPHEUS and ABS, which performed best in terms of testing effectiveness in RQ1, and compare them with BASFuzz across all datasets and threat models. Figure~\ref{Fig4} reports the time overhead required to generate a single successful test case. Given the significant differences in time costs between methods (ranging from hundreds to thousands of seconds), we use a logarithmic scale on the vertical axis for better visualization of the large-scale differences. The experimental results show that BASFuzz generates successful test cases in less time. Its average time overhead is reduced by 2163.852 seconds compared to MORPHEUS and by 160.469 seconds compared to ABS. This efficiency is attributed to BASFuzz's structural innovation in the fuzzing loop process, which combines beam search with simulated annealing. This approach significantly compresses redundant search paths while improving testing effectiveness. Additionally, BASFuzz introduces entropy-based beam width pruning to control the number of candidate variants per round dynamically, prioritizing high-value perturbation areas and effectively reducing invalid mutations.

\begin{figure}[t]
    \centering
    \includegraphics[width=1\textwidth, height=0.135\textheight]{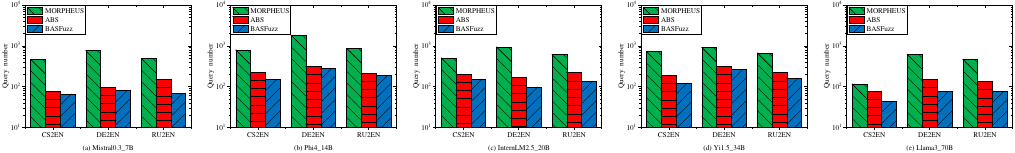}\\
    \caption{Results of test query number for different datasets and threat models (want $\downarrow$). }
    \label{Fig5}
\end{figure}

Figure~\ref{Fig5} shows the average number of queries made to the threat model during the testing process. For each successful test case generated, BASFuzz initiates 591.451 fewer inference calls than MORPHEUS, reducing the computational load on the underlying LLM during testing. MORPHEUS, which heavily relies on dense candidate sampling and large-scale morphological mutation strategies, is feasible in traditional DNN-based NLP software testing, as DNNs typically have fast inference speeds. However, when applied to large-scale LLMs with greater parameter sizes and longer response times, the inference cost is magnified, significantly reducing MORPHEUS's overall testing efficiency. The improvement in efficiency is especially important in industrial scenarios. LLM-based NLP software is often encapsulated through commercial APIs, and each inference call during testing incurs time delays and financial costs, such as cloud service charges based on the number of calls or characters processed. In sensitive industries that require frequent regression testing or security audits, excessive threat model calls quickly accumulate substantial costs. In contrast, BASFuzz, with its hybrid heuristic fuzzing loop, reduces redundant inference calls during the testing process, achieving higher robustness testing effectiveness with fewer software interactions.
\vspace{0.2cm}
\begin{mdframed}[backgroundcolor=gray!20, linecolor=black]
\textbf{Answer to RQ2:} 
BASFuzz, with its efficient fuzzing loop combining beam search, simulated annealing, and entropy-based beam width pruning, significantly reduces time overhead and query number while ensuring testing effectiveness. This lightweight characteristic makes BASFuzz more practically valuable in resource-constrained or large-scale testing scenarios.
\end{mdframed}
\subsection{RQ3: What is the contribution of different components to BASFuzz’s testing effectiveness?
}
\begin{table}[t]
\caption{Robustness testing results for Phi4\_14B using popular perturbation space construction methods, including synonym substitution with GloVe word embeddings (BASFuzz-G), synonym substitution with WordNet (BASFuzz-W), and related word substitution with OMW (BASFuzz-O).}
\label{tab3}
\begin{tabular}{c|c|cccccc}
\hline
Dataset & Method & S-rate & C-rate & PPL & G-error & TO & QN \\ \hline
 & BASFuzz-G & 61.168 & 4.634 & 189.478 & 15.066 & \textbf{223.599} & 209.773 \\
 & BASFuzz-W & 51.961 & 8.810 & 202.362 & 14.286 & 346.023 & 276.552 \\
 & BASFuzz-O & 58.829 & 5.991 & 182.361 & 13.748 & 325.697 & 239.495 \\
\multirow{-4}{*}{CS2EN} & BASFuzz & \cellcolor[HTML]{C0C0C0}\textbf{64.601} & \cellcolor[HTML]{C0C0C0}\textbf{3.553} & \cellcolor[HTML]{C0C0C0}\textbf{170.092} & \cellcolor[HTML]{C0C0C0}\textbf{12.029} & \cellcolor[HTML]{C0C0C0}284.207 & \cellcolor[HTML]{C0C0C0}\textbf{148.974} \\ \hline
 & BASFuzz-G & 67.988 & 4.992 & 129.077 & 15.853 & \textbf{362.758} & 311.942 \\
 & BASFuzz-W & 64.158 & 8.065 & 131.587 & 14.975 & 598.149 & 385.207 \\
 & BASFuzz-O & 70.687 & 6.089 & 127.587 & 14.254 & 587.463 & 331.036 \\
\multirow{-4}{*}{DE2EN} & BASFuzz & \cellcolor[HTML]{C0C0C0}\textbf{72.651} & \cellcolor[HTML]{C0C0C0}\textbf{4.743} & \cellcolor[HTML]{C0C0C0}\textbf{124.292} & \cellcolor[HTML]{C0C0C0}\textbf{12.618} & \cellcolor[HTML]{C0C0C0}553.291 & \cellcolor[HTML]{C0C0C0}\textbf{286.607} \\ \hline
 & BASFuzz-G & 63.123 & 4.462 & 66.459 & 14.105 & \textbf{221.173} & 218.674 \\
 & BASFuzz-W & 59.860 & 7.128 & 76.313 & 13.829 & 463.262 & 369.199 \\
 & BASFuzz-O & 64.222 & 4.236 & 69.313 & 13.113 & 435.882 & 335.341 \\
\multirow{-4}{*}{RU2EN} & BASFuzz & \cellcolor[HTML]{C0C0C0}\textbf{69.899} & \cellcolor[HTML]{C0C0C0}\textbf{3.147} & \cellcolor[HTML]{C0C0C0}\textbf{64.423} & \cellcolor[HTML]{C0C0C0}\textbf{12.727} & \cellcolor[HTML]{C0C0C0}350.431 & \cellcolor[HTML]{C0C0C0}\textbf{188.334} \\ \hline
\end{tabular}
\end{table}
To evaluate the effectiveness of perturbation space construction and fuzzing loop, we perform an ablation study on BASFuzz across three datasets. Table~\ref{tab3} shows the robustness testing results for Phi4\_14B using popular perturbation space construction methods. BASFuzz, combining the multilingual word network OMW with LLM-based word embeddings, achieves a success rate of 72.651\%, also demonstrating superior performance in test case stealthiness and fluency. BASFuzz-G, which uses GloVe for word substitution, benefits from the efficiency of preloaded word embedding space and distance calculation, enabling quick construction of candidate word sets, and exhibits the lowest time overhead. However, this simplified strategy falls behind BASFuzz in other evaluation metrics, particularly in test effectiveness and perturbation stealthiness. Test success rate, as a core metric of robustness testing ability, plays a decisive role in enhancing software robustness evaluation. Focusing solely on time efficiency while neglecting test effectiveness undermines the practical engineering value of the testing method. Moreover, although BASFuzz-G is the fastest in terms of testing speed, it queries the threat model more frequently than BASFuzz, with an average of 60.799 more queries per test case. The root cause lies in the static word vectors, which fail to accurately capture the semantic relationships between candidate words, resulting in a perturbation space filled with low-quality variants that require more attempts to generate effective perturbations. In contrast, BASFuzz uses LLM-based word embeddings, which excel at semantic understanding, to compute semantic consistency and further optimize the perturbation space, effectively improving testing efficiency. The comparison between BASFuzz-O and BASFuzz also shows that, although the two-stage perturbation space construction strategy introduces a semantic constraint step, the effective focus on high-quality candidate words greatly reduces subsequent invalid queries. This not only prevents an increase in overall time overhead due to encoding word vectors but also enhances test effectiveness while reducing query counts.

Table~\ref{tab4} presents the ablation results for the main submodules of the fuzzing loop. Based on standard beam search (\textit{w/o} SA\&EP), the method can achieve a certain level of global search and efficient perturbation, but is prone to local optima in high-dimensional and complex perturbation spaces, leading to poor test effectiveness and efficiency. After integrating the simulated annealing-based acceptance strategy into the fuzzing loop (\textit{w/o} EP), test effectiveness and perturbation stealthiness improve significantly, indicating that our simulated annealing mechanism provides a probabilistic acceptance channel for non-optimal perturbations, allowing the fuzzing process to maintain global exploration capability despite short-term score fluctuations. Upon introducing the entropy-based pruning strategy (\textit{w/o} SA), BASFuzz demonstrates even better efficiency in time overhead and query number. By calculating the distribution entropy of the current beam using BLEU scores, the beam width is dynamically adjusted to avoid redundant computations and inefficient paths. With the combination of these components, BASFuzz, in its complete configuration, demonstrates the best performance across all metrics on the three datasets, with test effectiveness improving by 7.852\%–9.869\%. The new fuzzing loop enables BASFuzz to effectively uncover robustness flaws while enhancing its usability in large-scale quality assurance processes, even under limited computational resources.
\begin{table}[t]
\caption{Ablation results of standard beam search (\textit{w/o} SA\&EP), selection via simulated annealing (\textit{w/o} EP), and entropy-based pruning (\textit{w/o} SA) on Phi4\_14B.}
\label{tab4}
\begin{tabular}{c|c|cccccc}
\hline
Dataset & Method & S-rate & C-rate & PPL & G-error & TO & QN \\ \hline
 & \textit{w/o} SA\&EP & 54.732 & 9.331 & 233.934 & 15.425 & 454.877 & 155.648 \\
 & \textit{w/o} EP & 57.656 & 6.722 & 204.914 & 14.282 & 391.283 & 176.196 \\
 & \textit{w/o} SA & 56.715 & 7.365 & 219.243 & 15.121 & 317.529 & 152.983 \\
\multirow{-4}{*}{CS2EN} & BASFuzz & \cellcolor[HTML]{C0C0C0}\textbf{64.601} & \cellcolor[HTML]{C0C0C0}\textbf{3.553} & \cellcolor[HTML]{C0C0C0}\textbf{170.092} & \cellcolor[HTML]{C0C0C0}\textbf{12.029} & \cellcolor[HTML]{C0C0C0}\textbf{284.207} & \cellcolor[HTML]{C0C0C0}\textbf{148.974} \\ \hline
 & \textit{w/o} SA\&EP & 64.125 & 7.145 & 156.587 & 14.844 & 649.448 & 303.617 \\
 & \textit{w/o} EP & 69.049 & 5.643 & 141.229 & 13.534 & 656.024 & 321.804 \\
 & \textit{w/o} SA & 65.705 & 6.669 & 150.143 & 14.556 & 598.263 & 291.452 \\
\multirow{-4}{*}{DE2EN} & BASFuzz & \cellcolor[HTML]{C0C0C0}\textbf{72.651} & \cellcolor[HTML]{C0C0C0}\textbf{4.743} & \cellcolor[HTML]{C0C0C0}\textbf{124.292} & \cellcolor[HTML]{C0C0C0}\textbf{12.618} & \cellcolor[HTML]{C0C0C0}\textbf{553.291} & \cellcolor[HTML]{C0C0C0}\textbf{286.607} \\ \hline
 & \textit{w/o} SA\&EP & 62.047 & 7.956 & 67.518 & 14.072 & 414.546 & 190.662 \\
 & \textit{w/o} EP & 65.464 & 5.097 & 65.425 & 12.932 & 439.074 & 221.421 \\
 & \textit{w/o} SA & 63.287 & 7.123 & 66.133 & 13.440 & 364.937 & 189.793 \\
\multirow{-4}{*}{RU2EN} & BASFuzz & \cellcolor[HTML]{C0C0C0}\textbf{69.899} & \cellcolor[HTML]{C0C0C0}\textbf{3.147} & \cellcolor[HTML]{C0C0C0}\textbf{64.423} & \cellcolor[HTML]{C0C0C0}\textbf{12.727} & \cellcolor[HTML]{C0C0C0}\textbf{350.431} & \cellcolor[HTML]{C0C0C0}\textbf{188.334} \\ \hline
\end{tabular}
\end{table}
\vspace{0.2cm}
\begin{mdframed}[backgroundcolor=gray!20, linecolor=black]
\textbf{Answer to RQ3:} 
Through the two-stage perturbation space construction strategy and modular fuzzing loop design of BASFuzz, not only improves testing effectiveness and efficiency, but also enhances the text quality of the generated test cases.
\end{mdframed}

\subsection{RQ4: How transferable are BASFuzz-generated test cases across different threat models?
}
In robustness testing, the transferability of test cases is an important metric for evaluating their broad applicability. We select the three baselines with the best current testing effectiveness and compare them with the test cases generated by BASFuzz across three datasets. Table~\ref{tab5} presents the transferability results across different threat models, where ``14b→70b'' indicates the transfer of test cases generated for Phi4\_14B to the testing effectiveness on Llama3\_70B, and vice versa. The test cases generated by BASFuzz maintain a high success rate in both settings, demonstrating better transferability than the baselines. In the scenario of transferring test cases from Phi4\_14B to Llama3\_70B, BASFuzz achieves a success rate of 87.101\%, which is 4.016\% higher than MORPHEUS's success rate when tested directly on Llama3\_70B. This suggests that the test cases generated by BASFuzz have broader applicability across different LLM-based software and can cover common robustness flaws present in various threat models. BASFuzz's perturbation space construction not only relies on static word substitution but also incorporates semantic constraints through LLM-based word embeddings, which include contextual knowledge. This ensures that the test cases generated by BASFuzz maintain semantic consistency across multiple threat models, avoiding over-reliance on training biases from a single model or overfitting to specific language patterns. Higher transferability offers better engineering value, reducing the workload and cost of testing each software separately, and ensuring long-term maintenance, especially as LLM-based software frequently updates. BASFuzz can efficiently and reliably perform robustness evaluations in such environments.

We also observe that the success rate of transferring test cases generated for Llama3\_70B to Phi4\_14B is typically lower than the reverse. Although Llama3\_70B has a larger parameter scale than Phi4\_14B, its older design and architecture result in less effective test cases when transferred to Phi4\_14B compared to test cases generated from Phi4\_14B. As a more recent model, Phi4\_14B has been optimized in terms of language modeling and training strategies, better adapting to complex language patterns and subtle perturbations, thus demonstrating greater robustness with test cases transferred from Llama3\_70B. This difference reflects the varying adaptability of different threat models when handling perturbations, further demonstrating that BASFuzz generates highly transferable test cases while considering the specific characteristics of LLM-based software.
\begin{table}[t]
\caption{The success rates of transferred adversarial test cases on the three datasets (want $\uparrow$).}
\label{tab5}
\begin{tabular}{c|c|cccc}
\hline
Dataset & Transfer relation & MORPHEUS & ABS & ABFS & BASFuzz \\ \hline
 & 14B→70B & 81.954 & 87.183 & 88.275 & \cellcolor[HTML]{C0C0C0}\textbf{89.967} \\
\multirow{-2}{*}{CS2EN} & 70B→14b & 42.857 & 51.709 & 50.602 & \cellcolor[HTML]{C0C0C0}\textbf{66.063} \\ \hline
 & 14B→70B & 73.334 & 76.538 & 75.667 & \cellcolor[HTML]{C0C0C0}\textbf{83.902} \\
\multirow{-2}{*}{DE2EN} & 70B→14b & 30.321 & 33.333 & 34.259 & \cellcolor[HTML]{C0C0C0}\textbf{50.704} \\ \hline
 & 14B→70B & 85.121 & 84.557 & 81.665 & \cellcolor[HTML]{C0C0C0}\textbf{87.435} \\
\multirow{-2}{*}{RU2EN} & 70B→14b & 41.189 & 42.718 & 44.569 & \cellcolor[HTML]{C0C0C0}\textbf{49.087} \\ \hline
\end{tabular}
\end{table}
\vspace{0.2cm}
\begin{mdframed}[backgroundcolor=gray!20, linecolor=black]
\textbf{Answer to RQ4:} 
The test cases generated by BASFuzz exhibit higher transferability compared to other baselines, reducing the time and resources required for testing each software individually, thereby enhancing the broad applicability and long-term value of the testing process.
\end{mdframed}

\subsection{RQ5: Can BASFuzz maintain robustness testing effectiveness when applied to the NLU task?}\label{sec6_5}
To validate the scalability of BASFuzz across different NLP task types, we apply it to text classification, a downstream task in NLU. To this end, we design and open-source a mutation-based fuzz testing tool based on BASFuzz, which supports full automation of the process, from objective function configuration and input perturbation generation to search process control. The tool, through a unified interface and highly configurable design, overcomes the current limitation of existing testing tools that are bound to specific task types, allowing testing methods originally designed for NLG tasks to be applied to NLU tasks and vice versa. Specifically, we use the confidence of the original labels in classification tasks as the objective function. If the generated variant causes the software to flip the input label, it is considered a successful test case. This setup not only retains the numerical optimization goal form consistent with NLG tasks but also maintains a consistent objective function semantics across multi-task testing, making the fuzz testing more efficient across different tasks.

Table~\ref{tab6} shows the comparison results of BASFuzz with five recent or widely used baselines. BASFuzz achieves the best testing effectiveness across all datasets and threat models, with an average success rate of 78.331\%, outperforming the classic DNN-based NLP software testing method TextFooler, which achieves a success rate of 67.912\%. BASFuzz's two-stage perturbation space construction improves the relevance and mutation quality of candidate words, ensuring that perturbations in NLU tasks maintain high semantic consistency. Furthermore, the collaborative design of beam search and simulated annealing in the fuzzing loop enhances the global search ability, thereby generating more effective test cases. In terms of the quality of the generated text, BASFuzz also performs excellently, with an average change rate of only 1.085 and an average perplexity of 51.347, outperforming other baselines in most cases. This indicates that the perturbations introduced by BASFuzz in NLU tasks are similarly hard to detect. 
We also observe cases where BASFuzz does not surpass the best-performing baselines. On certain models of the AG’s News and MR datasets, ABFS shows a slight advantage in change rate, while TextFooler achieves lower perplexity in a few cases. These results mainly stem from the fact that such baselines adopt more aggressive perturbation strategies or stricter text quality constraints, thereby gaining certain advantages in preserving fluency or minimizing perturbations, though often at the expense of testing effectiveness. Nevertheless, the magnitude of these differences is limited and does not undermine the significant advantages of BASFuzz in overall testing effectiveness and text quality. The testing strategy of BASFuzz exhibits good scalability across diversified inputs and task settings,
making it highly significant for the broad application of current LLM-based NLP software in multi-task and multi-domain scenarios.
\begin{table}[t]
\setlength{\tabcolsep}{1.4pt} 
\footnotesize
\caption{Comparison of the quality of test cases generated by six testing methods for the NLU task.}
\label{tab6}
\begin{tabular}{c|c|ccc|ccc|ccc|ccc|ccc}
\hline
 &  & \multicolumn{3}{c|}{Mistral0.3\_7B} & \multicolumn{3}{c|}{Phi4\_14B} & \multicolumn{3}{c|}{InternLM2.5\_20B} & \multicolumn{3}{c|}{Yi1.5\_34B} & \multicolumn{3}{c}{Llama3\_70B} \\ \cline{3-17} 
\multirow{-2}{*}{Dataset} & \multirow{-2}{*}{Baseline} & S-rate & C-rate & PPL & S-rate & C-rate & PPL & S-rate & C-rate & PPL & S-rate & C-rate & PPL & S-rate & C-rate & PPL \\ \hline
 & ABS & 66.173 & 1.756 & 52.771 & 88.894 & 0.987 & 47.542 & 77.058 & 1.283 & 49.102 & 75.305 & 1.663 & 52.641 & 62.411 & 1.316 & 48.548 \\
 & ABFS & 65.124 & 0.996 & 50.899 & 88.857 & 0.976 & 47.412 & 74.254 & 1.127 & 48.448 & 72.016 & 1.295 & 51.714 & 61.759 & 1.124 & 48.272 \\
 & GreedyFuzz & 51.597 & 1.885 & 57.342 & 33.728 & 1.898 & 58.703 & 73.952 & 1.943 & 56.151 & 21.073 & 1.971 & 56.542 & 12.761 & 1.763 & 57.669 \\
 & TextFooler & 62.949 & 1.095 & 51.363 & 79.492 & 1.541 & 52.822 & 73.345 & 1.251 & 50.363 & 73.232 & 1.881 & 52.937 & 59.499 & 1.168 & 50.749 \\
 & PWWS & 61.565 & 1.715 & 52.101 & 73.831 & 1.043 & 50.459 & 74.417 & 1.211 & 50.319 & 69.889 & 2.203 & 54.967 & 61.437 & 1.242 & 52.594 \\
\multirow{-6}{*}{FP} & BASFuzz & \cellcolor[HTML]{C0C0C0}\textbf{72.041*} & \cellcolor[HTML]{C0C0C0}\textbf{0.961*} & \cellcolor[HTML]{C0C0C0}\textbf{49.139*} & \cellcolor[HTML]{C0C0C0}\textbf{91.149*} & \cellcolor[HTML]{C0C0C0}\textbf{0.942*} & \cellcolor[HTML]{C0C0C0}\textbf{46.954*} & \cellcolor[HTML]{C0C0C0}\textbf{83.118*} & \cellcolor[HTML]{C0C0C0}\textbf{0.936*} & \cellcolor[HTML]{C0C0C0}\textbf{45.169*} & \cellcolor[HTML]{C0C0C0}\textbf{77.078*} & \cellcolor[HTML]{C0C0C0}\textbf{1.225*} & \cellcolor[HTML]{C0C0C0}\textbf{51.334*} & \cellcolor[HTML]{C0C0C0}\textbf{65.177*} & \cellcolor[HTML]{C0C0C0}\textbf{0.949*} & \cellcolor[HTML]{C0C0C0}\textbf{47.755*} \\ \hline
 & ABS & 66.934 & 1.197 & 46.905 & 73.083 & 1.002 & 46.439 & 72.026 & 1.199 & 48.148 & 75.943 & 1.641 & 49.919 & 68.369 & 1.407 & 46.898 \\
 & ABFS & 67.912 & 0.885 & 46.363 & 72.736 & 0.921 & 46.098 & 68.574 & 1.129 & 46.103 & 76.777 & \textbf{0.999} & \textbf{46.797} & 62.682 & 1.093 & 46.507 \\
 & GreedyFuzz & 48.999 & 1.477 & 52.626 & 18.978 & 1.443 & 53.364 & 50.168 & 1.511 & 51.886 & 2.212 & 1.046 & 58.035 & 12.151 & 1.504 & 49.053 \\
 & TextFooler & 66.445 & 1.479 & 47.254 & 43.664 & 1.378 & 46.992 & 71.562 & 1.849 & 48.378 & 74.403 & 2.439 & 51.546 & 59.481 & 1.648 & 48.141 \\
 & PWWS & 68.488 & 0.892 & 47.759 & 41.845 & 0.933 & 47.538 & 71.633 & 1.512 & 48.001 & 75.292 & 1.054 & 46.892 & 61.474 & 1.865 & 48.594 \\
\multirow{-6}{*}{\begin{tabular}[c]{@{}c@{}}AG's\\ News\end{tabular}} & BASFuzz & \cellcolor[HTML]{C0C0C0}\textbf{73.025*} & \cellcolor[HTML]{C0C0C0}\textbf{0.818*} & \cellcolor[HTML]{C0C0C0}\textbf{45.351*} & \cellcolor[HTML]{C0C0C0}\textbf{76.055*} & \cellcolor[HTML]{C0C0C0}\textbf{0.898*} & \cellcolor[HTML]{C0C0C0}\textbf{45.326*} & \cellcolor[HTML]{C0C0C0}\textbf{74.098*} & \cellcolor[HTML]{C0C0C0}\textbf{0.868*} & \cellcolor[HTML]{C0C0C0}\textbf{44.163*} & \cellcolor[HTML]{C0C0C0}\textbf{78.176*} & \cellcolor[HTML]{C0C0C0}2.161 & \cellcolor[HTML]{C0C0C0}50.506 & \cellcolor[HTML]{C0C0C0}\textbf{75.538*} & \cellcolor[HTML]{C0C0C0}\textbf{0.889*} & \cellcolor[HTML]{C0C0C0}\textbf{45.425*} \\ \hline
 & ABS & 48.915 & 2.799 & 73.099 & 86.516 & 1.230 & 59.391 & 83.086 & 1.666 & 61.869 & 83.786 & 1.867 & 63.589 & 71.336 & 1.954 & 61.411 \\
 & ABFS & 46.855 & \textbf{1.265} & 63.211 & 84.862 & 1.176 & 58.755 & 79.318 & 1.502 & 58.121 & 78.927 & 1.218 & 60.271 & 68.017 & 1.168 & 58.178 \\
 & GreedyFuzz & 28.981 & 1.892 & 67.262 & 27.872 & 1.919 & 65.722 & 61.645 & 1.941 & 66.622 & 10.782 & 1.947 & 68.162 & 41.315 & 1.922 & 68.574 \\
 & TextFooler & 58.418 & 1.287 & \textbf{61.287} & 70.795 & 1.529 & 60.915 & 82.211 & 1.913 & 62.139 & 80.747 & 1.945 & 63.623 & 62.433 & 2.053 & 62.779 \\
 & PWWS & 47.815 & 4.877 & 87.523 & 59.312 & 1.796 & 58.741 & 78.031 & 2.048 & 63.011 & 81.164 & 2.092 & 66.575 & 68.127 & 1.324 & 60.772 \\
\multirow{-6}{*}{MR} & BASFuzz & \cellcolor[HTML]{C0C0C0}\textbf{69.077*} & \cellcolor[HTML]{C0C0C0}1.282 & \cellcolor[HTML]{C0C0C0}62.998 & \cellcolor[HTML]{C0C0C0}\textbf{89.158*} & \cellcolor[HTML]{C0C0C0}\textbf{1.099*} & \cellcolor[HTML]{C0C0C0}\textbf{57.094*} & \cellcolor[HTML]{C0C0C0}\textbf{88.066*} & \cellcolor[HTML]{C0C0C0}\textbf{1.023*} & \cellcolor[HTML]{C0C0C0}\textbf{61.397*} & \cellcolor[HTML]{C0C0C0}\textbf{87.169*} & \cellcolor[HTML]{C0C0C0}\textbf{1.105*} & \cellcolor[HTML]{C0C0C0}\textbf{59.859*} & \cellcolor[HTML]{C0C0C0}\textbf{76.044*} & \cellcolor[HTML]{C0C0C0}\textbf{1.117*} & \cellcolor[HTML]{C0C0C0}\textbf{57.729*} \\ \hline
\end{tabular}
\end{table}

\begin{mdframed}[backgroundcolor=gray!20, linecolor=black]
\textbf{Answer to RQ5:} 
With the configurable objective function and fuzzing technique of BASFuzz, it achieves leading testing effectiveness and perturbation stealthiness in NLU tasks, such as text classification. This result fully demonstrates the scalability of BASFuzz across multiple NLP task scenarios, laying the foundation for generalized fuzz testing of LLM-based NLP software.
\end{mdframed}

\section{Threats to Validity}\label{sec7}
\subsection{Internal Threats}
The internal validity of BASFuzz is primarily influenced by its supporting components, including the multilingual word network, LLM-based word embeddings, beam search, and simulated annealing. While these modules have been widely adopted and empirically validated in their respective domains, the fuzzing workflow they collectively form can only approximate global optima. It cannot guarantee the generation of absolutely optimal test cases in every trial, which is a common limitation shared by all heuristic search–based methods. To mitigate the impact of these components, we conduct a set of experiments on testing effectiveness, perturbation stealthiness, and testing efficiency, comparing BASFuzz against representative baselines. BASFuzz consistently outperforms the baselines across all key metrics, suggesting that its advantage lies not only in component synergy but also in its methodological innovations. In addition, hyperparameter settings may also pose threats to internal validity—for instance, beam width range and cooling factor in the fuzzing loop. To control such bias, we strictly follow original configurations reported in the literature when reproducing baselines, ensuring experimental fairness and reproducibility. For BASFuzz, we perform parameter tuning and apply consistent hyperparameter configurations across all datasets and threat models, ensuring cross-setting comparability. To further eliminate potential sources of error, we conduct a validation of the experimental code and procedures. All experimental artifacts, including code and data, are made publicly available in a reproducible repository~\cite{BASFuzz} to support independent verification and further research by the community.

\subsection{External Threats}
The external validity of BASFuzz is mainly constrained by dataset coverage and method generalizability. Our evaluation focuses on multilingual machine translation tasks within the Indo-European language family. These languages typically feature rich morphological systems, which may not fully represent challenges posed by typologically distant or logographic languages. Nevertheless, BASFuzz leverages OMW, which natively supports dozens of natural languages and exhibits strong multilingual morphological compatibility. Additionally, the LLM-based word embeddings incorporated into BASFuzz enable context-aware semantic encoding across languages, providing adaptability and scalability in multilingual settings. In our experiments, we construct source inputs from three language pairs and evaluate across five threat models. BASFuzz achieves the highest testing effectiveness in all cases, further validating its cross-lingual generalizability. Although the main experiments center on NLG tasks, the fuzzing strategy of BASFuzz does not rely on task-specific textual features. Instead, it focuses on generic perturbation and search mechanism. By adapting the objective function for different downstream NLP tasks, BASFuzz can be readily transferred to other application scenarios. In RQ5, we extend BASFuzz to text classification as a representative NLU task, demonstrating its scalability and laying a solid foundation for broader applications in LLM-based NLP software testing.

\section{Related Work}\label{sec8}
With the widespread integration of LLMs into NLP software, the research community increasingly recognizes the potential risks posed by robustness issues~\cite{10.1145/3719006,10.1145/3660808}. In both security-critical domains such as finance~\cite{10.1145/3688399}, healthcare~\cite{qiu2024llm}, and public-facing applications like dialogue and generative systems~\cite{10.1145/3664608,10.1145/3730578}, robustness defects may lead to severe safety hazards. As a testing technique that does not require white-box access, fuzz testing has been widely adopted in recent years for robustness evaluation of intelligent software~\cite{xiao2024assessing,xiao2025abfs,10.24963/ijcai.2024/730,Cheng_Yi_Chen_Zhang_Hsieh_2020,tan-etal-2020-morphin,Jin_Jin_Zhou_Szolovits_2020,ren-etal-2019-generating}. Its primary objective is to generate input variants that can expose robustness vulnerabilities. In robustness evaluation for LLM-based NLP software, the stronger resilience and longer inference time of LLMs compared to DNNs make manual construction of test cases inefficient and costly. Therefore, this work focuses on automated fuzzing techniques, which can be classified into three classes~\cite{10.1145/3597503.3639121}: generation-based fuzzers, learning-based fuzzers, and mutation-based fuzzers.
\subsection{Generation-based Fuzzer}
Generation-based fuzzers are a traditional approach in automated software testing. They typically rely on predefined mutation rules or generation templates to systematically produce test cases that cover diverse regions of the input space. This paradigm has also been widely applied to robustness testing in NLP software. In the context of content moderation systems, Wang et al.~\cite{10172598} proposed a taxonomy of 11 transformation strategies spanning character, word, and sentence levels to generate test cases automatically. For named entity recognition, Yu et al.~\cite{10.1145/3611643.3616295} introduced TIN, a testing method that constructs semantically similar yet syntactically altered variants via paraphrasing, structural transformation, and entity shuffling. Another common practice is to construct offline datasets that contain a large number of inputs with potential robustness flaws, forming static testing benchmarks. Wang et al.~\cite{wang2024robustness} built robustness testing datasets based on AdvGLUE and ANLI by injecting word-level, sentence-level, and manually crafted perturbations. Yuan et al.~\cite{NEURIPS2023_b6b5f50a} proposed BOSS, which covers five major NLP tasks, each with one in-distribution dataset and three out-of-distribution robustness datasets. However, static dataset construction raises the risk of data leakage. If the threat model is exposed to similar inputs during training, the fairness and credibility of testing results can be compromised. Moreover, as LLM-based software evolves rapidly, static test sets struggle to align with shifting software semantics and behavior, leading to reduced coverage and representativeness. To overcome these limitations, BASFuzz avoids static generation strategies. Instead, it constructs test cases dynamically through real-time interaction with the threat model, making it better suited to the evolving nature and agile development requirements of LLM-based NLP software.
\subsection{Learning-based Fuzzer}
To address the limitations of traditional fuzzing in complex task settings and emerging software types, researchers have recently proposed learning-based fuzzers. These approaches aim to incorporate neural models into the testing process to generate more adaptive test cases automatically. Yao et al.~\cite{10.1145/3688839} employed the policy gradient method from reinforcement learning, designing a reward function based on code-text consistency, fluency, and alienation rate to generate minimally perturbed inputs. For mobile applications, Liu et al.~\cite{10.1145/3597503.3639118} utilized LLMs to generate text inputs that can pass graphical user interface-based page validation and then leveraged LLMs again to create mutation rules for constructing fuzzing generators. Ugarte et al.~\cite{10.1145/3713081.3731733} defined a jailbreak robustness coverage matrix from three perspectives—safety category, writing style, and persuasive technique—using OpenAI API assistants and retrieval-augmented generation to create test cases. While learning-based fuzzers improve automation and adaptability, their effectiveness remains constrained by the capacity of their underlying models. When the threat model possesses significantly stronger representational power than the model used for generation, the latter often fails to produce effective test cases, resulting in conservative robustness assessments. Furthermore, learning-based fuzzers suffer from slow convergence and poor generalization, which undermines testing efficiency. BASFuzz does not rely solely on neural generation. Instead, it combines multilingual lexical resources and a dynamic fuzzing loop to explore the perturbation space heuristically. This design avoids the performance bottlenecks inherent in learning-based fuzzers.
\subsection{Mutation-based Fuzzer}
Mutation-based fuzzers iteratively apply transformations to seed inputs to generate new mutated variants. Earlier mutation-based techniques mainly focused on two input components: prompts and examples, applying perturbations to evaluate robustness. Zhang et al.~\cite{10.1145/3691620.3695001} analyzed the software architecture of LLM-based systems to identify sensitive features, crafting test cases by modifying key information or distorting prompt semantics. For chatbot-based web applications, Pedro et al.~\cite{11029790} evaluated jailbreak robustness using both direct malicious prompt injection via web interfaces and stored malicious content in user-generated data. These methods target prompt injection to trigger prohibited behavior, such as generating unethical content. In contrast, BASFuzz focuses on assessing the software's decision consistency under fine-grained perturbations. This fundamental difference in testing objectives and impact scope distinguishes BASFuzz from prior methods. In an example-oriented robustness evaluation, Huang et al.~\cite{10.1145/3688835} extracted semantic features from each neuron of the threat model, constructing natural language descriptions linked to neural functionalities and applying transformations like pixel-level noise, affine modifications, and style transfer to enrich test diversity. Xiao et al.~\cite{10298415} used WordNet to construct the perturbation space and employed adaptive particle swarm optimization with Lévy flights to enhance fuzzing efficiency on DNN-based NLP software. Notably, current mutation-based fuzzers often neglect the complex interaction between prompts and examples during real-world software execution. Since the output of LLM-based NLP software is jointly influenced by both components, fuzzing either in isolation fails to expose robustness flaws fully and can lead to false conclusions. To address this, BASFuzz tests the entire input, capturing its interaction to more accurately evaluate software robustness under joint input perturbation.

\section{Conclusion}\label{sec9}
In this paper, we present BASFuzz, a mutation-based fuzzing method designed for LLM-based NLP software. BASFuzz targets the full input, composed of both prompts and examples. It constructs a semantics-preserving perturbation space using multilingual lexical resources and LLM-based word embeddings. BASFuzz employs a fuzzing loop driven jointly by beam search and simulated annealing to evaluate adversarial robustness efficiently. We conduct extensive experiments across six real-world datasets and five threat models. Results show that BASFuzz achieves a 5.338\% improvement in average testing effectiveness compared to the strongest existing baselines, while reducing the time cost per successful test case by 82.984\%. This demonstrates that BASFuzz enables more effective robustness evaluations with significantly lower testing overhead. Furthermore, the generated test cases exhibit superior stealth and transferability. In future work, we plan to enhance BASFuzz's applicability to commercial closed-source software and investigate its adaptability in multilingual and multimodal scenarios to strengthen its practical utility further.

\begin{acks}
The work is supported by the National Natural Science Foundation of China (62272145, U21B2016), and the Fundamental Research Funds for the Central Universities (B240205001).
\end{acks}

\bibliographystyle{ACM-Reference-Format}
\bibliography{ref}

\end{document}